\documentclass[%
 reprint,
nofootinbib,
 amsmath,amssymb,
 aps,
]{revtex4-2}

\usepackage{graphicx}
\usepackage{dcolumn}
\usepackage{bm}
\usepackage[capitalise]{cleveref}
\usepackage{xcolor}
\usepackage{amsmath}
\usepackage{subcaption}
\begin{document}
\newcommand{\red}[1]{\textcolor{red}{#1}}

\preprint{APS/123-QED}

\title{Data-Driven Discovery of Beam Centroid Dynamics}

\author{Liam A. Pocher}
\email{lpocher@umd.edu}
\author{Irving Haber}%
\author{Thomas M. Antonsen Jr.}%
\author{Patrick G. O'Shea}
\affiliation{Institute for Research in Electronics and Applied Physics, University of Maryland, College Park, MD, USA}

\date{\today}

\begin{abstract}
Understanding and predicting complex dynamics in accelerators is necessary for their successful operation. A grand challenge in accelerator physics is to develop predictive virtual accelerators that mitigate design cost and schedule risk. Data-driven techniques greatly appeal to generating virtual accelerators due to their limited dimensionality compared with first-principle simulation, yet require significant up-front investment and lack interpretability in the context of governing equations. This paper uses an alternative, interpretable, data-driven technique called Sparse Identification of Nonlinear Dynamics (SINDy) developed by University of Washington researchers to study nonlinear beam centroid dynamics excited by realistic beam injection. We propose evolution equations based solely on data analysis and intuition of the underlying lattice structure, without recourse to an underlying first-principles centroid model nor the actual lattice forcing functions. We do this to mimic an application environment where analytic models are inadequate or where detailed lattice forcing functions are unknown. In the context of the accurate centroid model, we report and interpret SINDy's beam evolution equations learned from the training data and show favorable prediction results. We compare with an alternative machine learning model used on the same training data and contrast its prediction ability, computational expense, and interpretability with SINDy's results.
\end{abstract}

\maketitle


\section{\label{sec:level1}Introduction}

Particle accelerators are complex and expensive. Understanding and predicting their beam dynamics is necessary for their successful operation. To address this, the Department of Energy (DOE) Office of High Energy Physics (HEP) has publicized a set of grand challenges that are critical for future DOE-HEP programs~\cite{nagaitsev2021accelerator}. One of these grand challenges is to develop predictive virtual accelerator models, which can significantly reduce design costs and experimental planning risk. 

Virtual models such as first-principle simulations require ever-increasing computational resources for high-fidelity simulations, validated to experiment and verified to theory. Even with simulations providing complete beam diagnostics not realizable in experiments, simulations are still cost-prohibitive because of the high dimensional nature of beam particle motion in accelerators. 
Further, due to the many focusing elements in an accelerator, optimization of these elements requires many simulations. 
Two classes of optimization have been applied in the field of accelerator physics, model-based and data-driven.
The computational burden of optimization is reduced to some degree by automatic differentiation~\cite{kaiser2024bridging}, user interface tools~\cite{paulson1995accelerator,huang2013algorithm}, and adjoint methods~\cite{dovlatyan2022optimization}. Adjoint approaches may be employed to reduce simulation parameter space, but still require user feedback on specific quantities to optimize. Recently, model free, data-driven methods have been introduced. These are Bayesian optimization~\cite{roussel2024bayesian}, reinforcement learning~\cite{kain2020sample,kaiser2022learning}, and machine learning (ML)~\cite{edelen2020machine,scheinker2021adaptive,boehnlein2022colloquium,roussel2023phase,roussel2024efficient}. These methods allow one to predict complex beam dynamics with high fidelity but require accurate training data, necessitating significant up-front investment. Also, these methods lack interpretability concerning the beam's evolution equations. 

In contrast, an alternative data-driven technique from the nonlinear dynamics community called Sparse Identification of Nonlinear Dynamics (SINDy)~\cite{de2020pysindy,Kaptanoglu2022} allows one to project simulation or experimental data onto a prescribed set of differential equations. The prescribed ordinary differential equations are populated with evolution terms (to-be-named basis functions) selected by the user based on likely underlying dynamics. This approach is applicable in accelerator environments where analytic models are inadequate or where detailed lattice forcing functions are unknown. The optimum coefficients for these terms are directly identifiable with the data's dynamics, in contrast to other data-driven techniques, which may be more accurate but lack interpretability. In addition, our approach is different from the recently developed adjoint method~\cite{dovlatyan2022optimization}. That method can be used for the design and optimization of lattices, whereas this method is more readily applicable to predicting long-term behavior of underlying beam dynamics. The technique we use bridges the gap between purely model-based and purely data-driven optimization.
 
We have presented prior work with SINDy applied to accelerator physics~\cite{pocher:napac22-tuze3,pocher:ipac23-thpl036}. In this paper, we study an example problem of the nonlinear dynamics of transverse centroid motion arising from realistic beam injection in the University of Maryland Electron Ring (UMER)~\cite{oshea:pac01-toaa008,kishek2014university}. We simulate beam dynamics with the particle-in-cell code Warp~\cite{grote2005warp} because it is well-validated through experimental measurments on UMER behavior~\cite{haber2007measurement,haber2007scaled}. We study this problem for multiple reasons:
\begin{enumerate}
    \item Simulations provide data to study dynamics around the entire ring; experimental data for UMER is too sparse.
    \item We can compare with prior work where ML was used to predict transverse centroid motion as well~\cite{komkov2019reservoir}.
    \item We can leverage well-validated simulations for a reliable and reproducible machine, UMER, thus providing a real experiment to ground our results.
    \item The elements in UMER are well understood, so we can deploy a semi-analytic centroid model to verify our approach. Since this model has analytic terms describing centroid motion, we can readily interpret the learned coefficients of the evolution terms we prescribe to SINDy a posteriori.
\end{enumerate} 
We prescribe evolution terms for SINDy solely by performing data analysis of the Warp simulated centroid motion, ignoring terms coincidentally duplicated in the centroid model.

We organize the paper as follows. Section~\ref{sec:methods} details the UMER accelerator and Warp simulations, provides an overview of SINDy, and motivates basis function prescription. Section~\ref{sec:res} describes results obtained for various SINDy fits with progressively more terms based on data analysis yielding higher fidelity fits. This section compares various SINDy fits varying optimization hyperparameters before detailing stable prediction. Section~\ref{sec:centroid} then introduces the centroid model and compares it to simulation data, providing verification and allowing us to interpret fit results. Section~\ref{sec:allcomp} compares and contrasts one SINDy fit to the centroid model and simulation data. Finally, Sec.~\ref{sec:disc} provides a paper overview, and possible extensions, and synthesizes outcomes.

\section{\label{sec:methods}Methods}

\subsection{\label{sec:method:UMER_lat}Accelerator Lattice}

\begin{figure}[!htbp]
  \centering
  \includegraphics[width=\columnwidth]{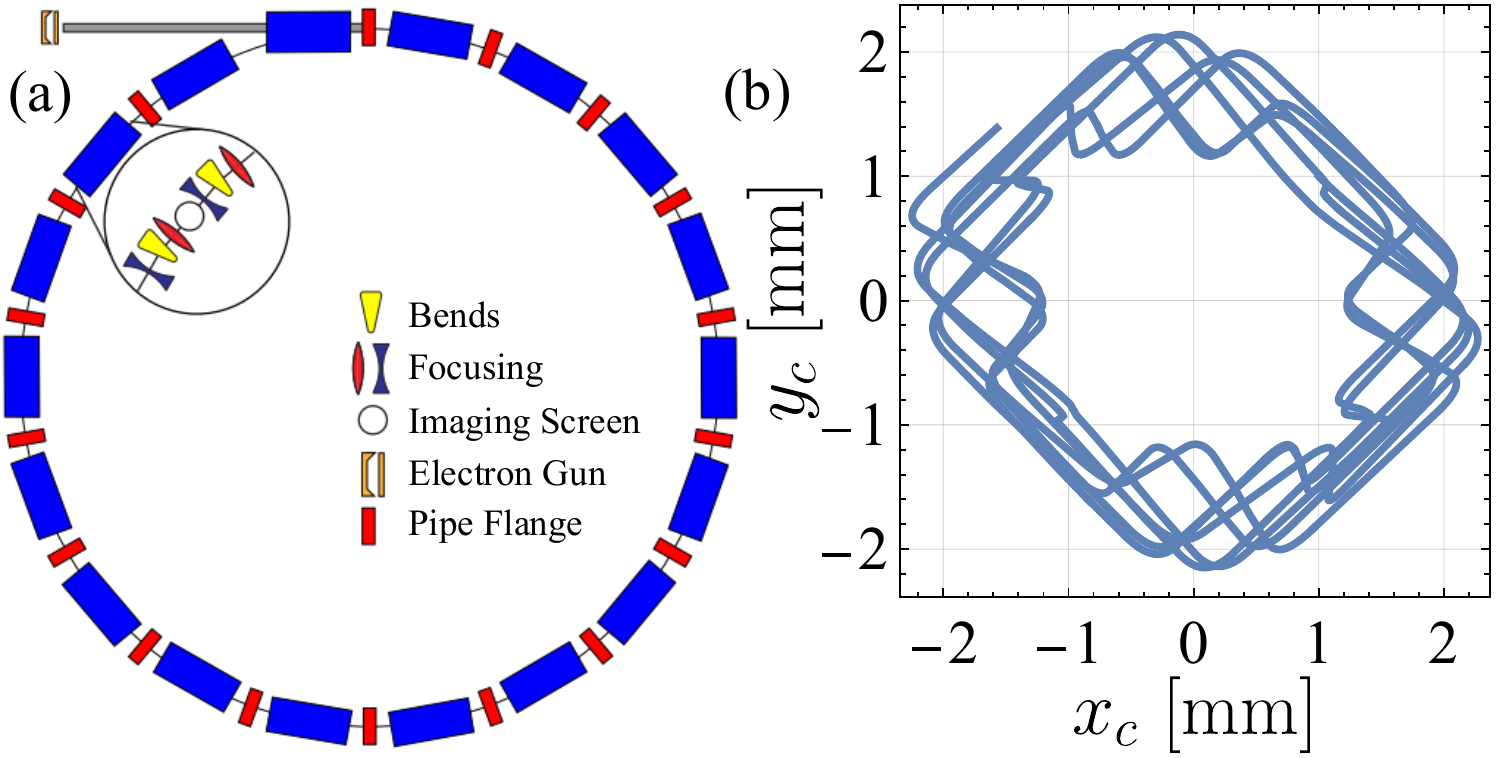}
\caption{\label{fig:UMER_lat_phase}(a) A top-down view diagram of UMER~\cite{oshea:pac01-toaa008,kishek2014university} reproduced from Komkov et al~\cite{komkov2019reservoir}. (b)  Beam centroid data from Warp calculated every 5~mm plotted in parametric phase space with respect to pipe center for one turn. The horizontal centroid is defined as $x_c$, see \cref{eqn:centroid_xc}, and the vertical centroid is defined as $y_c$, see \cref{eqn:centroid_yc}.}
\end{figure}

We employ a simulation inspired by UMER, \cref{fig:UMER_lat_phase}(a), in Warp that has been well-validated on the UMER ~\cite{haber2007scaled,haber2007measurement,kishek2003simulations}. UMER is a rounded-vertex, 36-sided polygon whose straight edges correspond to focus-drift-defocus-drift (FODO) quadrupoles\footnote{Defocusing and focusing for quadrupoles refers to the effect they have on the $\hat{x}$ beam envelope. The (de)focus descriptor is switched for $\hat{y}$.}. Between adjacent straight sections are dipole bends giving UMER its rounded vertex appearance as shown in \cref{fig:UMER_lat_phase}(a). UMER is a scaled down accelerator lattice, roughly 3.6~m in diameter, operating with a 10~keV electron beam of variable current up to  100~mA. The basic parameters of UMER are given in \cref{tab:umer_char}.

\begin{table}[!htbp]
	\setlength\tabcolsep{3.5pt}
	\centering
	\caption{\label{tab:umer_char}UMER lattice parameters used in this paper.}
 \begin{ruledtabular}
	\begin{tabular}{lr}
		Parameter & Value \\ \hline
            Current Range $I$ & 600~{\textmu}A \\
            Tune & 6.74 \\
            Periods & 36  \\ 
            Period Length & 0.32~m \\
            Superperiod Length & 11.52~m \\
		Undepressed Phase Advance & 76$^{\circ}$ per 0.32~m \\
		Defocusing Quad. Strength & 4.63~Gs/cm\\
            Defocusing Quad. Length & 5.64~cm\\
            Focusing Quad. Strength & -4.64~Gs/cm \\
            Focusing Quad. Length & 5.42~cm\\
            First Quadrupole Center &  8~cm  \\
            Dipole Strengths & \cref{tab:model_dipole_by} in Appendix~\ref{sec:app_dipo} \\ 
            Dipole Length &  3.6~cm \\
            First Dipole Center & 17.8~cm
	\end{tabular}
  \end{ruledtabular}
\end{table}

The beam we study in this paper is a low current UMER setting of 600~{\textmu}A, with a corresponding generalized perveance $K$ of 8.96~{\textmu}Perv. The beam radius---which is twice the variance, $\sigma_x$, in either of the transverse directions---is approximately 2~mm, with a rms-edge emittance of 7.6~{\textmu}m. The generalized perveance is defined as $K = 2 I \, I_0^{-1} \beta^{-3} \gamma^{-3}$, and is a measure of the beam's repulsive electrostatic effects to inertial effects. The rms-edge emittance is a measure of beam area in phase space, and often used a metric of beam quality; the smaller the beam's emittance the higher the beam's quality. These beam settings induce a dimensionless space-charge to emittance ratio of order unity, $K \sigma_x^2  \varepsilon_x^{-2}\approx 0.6$~\cite{reiser2008theory}.

The simulation uses the nominal UMER energy of 10~keV. The simulation pipe radius $r_p$ is 25.4~mm, matching UMER's physical pipe radius. There are 64 numerical cells across the beam across the $x$ and $y$ coordinates, with an axial resolution of 5~mm, and two hundred thousand numerical particles. The beam is initialized with a semi-Gaussian\footnote{The semi-Gaussian distribution is a uniformly filled ellipse in transverse configuration space and has a Gaussian distribution in transverse momentum space approximating thermal beam velocities.} trace space distribution with an initial offset of 1~mm from the beam pipe center in both transverse $x$ and $y$ directions with no angle offset ($x'=y'=0$~mrad), representing a realistically misinjected beam. The simulation does not include a measurement of the Earth's field. The simulation's beam tune is approximately 6.74, which is close to the zero-current tune of UMER.

Our studies focus on the transverse dynamics of the electron beam as it circulates in UMER. We apply SINDy to analyzing transverse centroid dynamics of an electron beam from the Warp simulation of UMER over one turn and then predicting over further turns. A reason we only use one turn for training is so we can make a direct comparison to work using ML~\cite{komkov2019reservoir}. A caveat to only using one turn is that the beam is statistically nonstationary on this timescale\footnote{We define the beam as statistically stationary if statistical properties of the beam are uniform in time, or unchanging on the timescale of the observation. A metric we used to calculate the stationarity of the beam was the horizontal centroid time the vertical centroid, a proxy measure of their correlation. Without dipoles (making UMER a linear accelerator, effectively) this quantity precesses about the center of the beam pipe on a period of two turns, while with dipoles (the ring configuration) this precession wavelength is increased an order of magnitude to 20 turns ($\sim$200~m). As we observe the beam in these current simulations on a timescale shorter than this, we describe the beam's centroid evolution as nonstationary.} so we cannot expect SINDy to reproduce secular behavior on timescales longer than this. 

The nonlinear dynamics\footnote{To verify that chaos wasn't present in UMER's digital representation, we performed an additional simulation with a $\pi/50$ radian angular offset and compared to \cref{fig:UMER_lat_phase}(b). That solution did not exhibit the typical sensitive dependence on initial conditions present in chaotic systems~\cite{ott2002chaos}.} governing UMER's centroid evolution cannot be solved analytically. However, the equations can be solved numerically from analytic models, producing semi-analytic results for centroid dynamics. In the simplest case, It is assumed that there is a closed trajectory in the ring that passes through the field nulls of the quadrupoles and is bent in the horizontal plane by the dipoles so that the trajectory is closed. The deviation of an individual particle’s coordinates from this reference closed trajectory is given by the two functions $x(z)$ and $y(z)$, where $x(z)$ measures outward radial displacement in the plane of the ring, $y(z)$ measures vertical displacement out of the plane of the ring, and $z$ is the distance along the reference trajectory. These deviations satisfy second-order differential equations in $z$.  Thus, the trajectory of an individual particle can be regarded as the evolution of a point in the four-dimensional phase space $(x, x’, y, y’ )$, where the prime denotes differentiation with respect to $z$.  For simplicity, we assume that all particles have the same velocity, $v_z=dz/dt$, along the reference trajectory.

When many particles are present, it is appropriate to consider averages of various moments of the coordinates.  In this paper, we will consider primarily the first moment, which defines the centroid of the beam
\begin{subequations}
\begin{align}
    x_c(z) &= \langle x \rangle \,, \label{eqn:centroid_xc} \\
    y_c(z) &= \langle y \rangle \,  .  \label{eqn:centroid_yc}
\end{align} 
\end{subequations}
Here, the angular brackets imply average over the ensemble of particles. See App.~\ref{sec:app_centroid} for further details on ensemble averaging. \Cref{fig:UMER_lat_phase}(b) shows the evolution of the beam centroid in the $x_c$ - $y_c$ plane for a beam simulated with the code WARP.  The motion of the centroid in this case consists of quasi-periodic oscillations superimposed on a spiraling motion about the reference trajectory, $x_c = y_c =0$.

In the paraxial limit, where the transverse oscillations of the beam are small,  $|x|$ , $|y| , \ll  R_s$ and $x’$, $y’ \ll 1$ where $R_s$ is the typical distance over which the focusing fields vary transverse to the reference trajectory, the evolution of the centroid is described by the pair of equations~\cite{reiser2008theory}
\begin{subequations}
\begin{align} 
    \label{eqn:xc_cent} x_c'' + \left[ \kappa_{x}(z) - \frac{1}{R^2(z)} \right] x_c &= 0 \, , 
    \\ 
    \label{eqn:yc_cent} y_c''  + \kappa_{y}(z) y_c &= 0 \, .
\end{align}
\end{subequations}

Here, $x_c / R^2(z)$ represents the change in the balance between the bending force of the dipoles and the centripetal acceleration implied by the curved trajectory.  If the beam is displaced radially such that $x_c > 0$, the centripetal acceleration decreases and the dipole force exceeds the centripetal acceleration, causing the beam to be directed back toward the reference trajectory.  Similarly, if the beam is radially displaced inward such that $x_c < 0$, the increased centripetal acceleration causes the beam to be directed outward toward the reference trajectory. Although this term represents a competition between two effects, centripetal acceleration and magnetic dipole force, it is represented in terms of a single variable related to an effective radius of curvature. 

The dipoles used in the simulation have strengths directly taken from a UMER experiment. However, since the simulation excludes the Earth's field, the dipole strengths do not fully bend the around into a closed loop after one lattice superperiod. Instead, the total revolution angle for one turn of the superperiod is 289$^{\circ}$, creating a ring whose closure length is 25\% larger than UMER's. To mimic UMER with higher fidelity--but not completely---the inclusion of a dipole strength correction accounting for the Earth's field in both the $\hat{x}$-direction and $\hat{y}$-direction would need to be included. We choose not to include these contributions to both reduce the learnable physics dataset, and also to use the same data as used in prior ML work~\cite{komkov2019reservoir}.

\begin{figure}[!htbp]
  \centering
  \includegraphics*[width=\columnwidth]{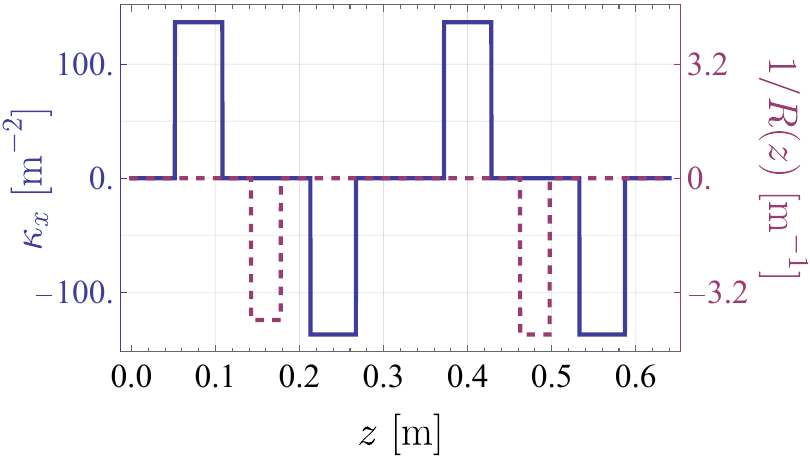}
\caption{\label{fig:kx_hard} Hard-edge profile of UMER lattice $\kappa_{x}(z)$, quadrupoles focusing the beam, in solid blue (left vertical axis) and $1/R(z)$, and dipoles bending the beam, in dashed purple (right vertical axis) used in the centroid model. Dipole strengths vary throughout the lattice and are not periodic on a period shorter than the superperiod of the ring, see \cref{tab:model_dipole_by}.}
\end{figure}

The terms $\kappa_x$ and $\kappa_y$ describe the effect of the quadrupoles.  If, as assumed here, the quadrupoles are properly aligned with respect to the normal to the plane of the ring, the quadrupole force is in the same direction $(+/-)$ as the displacement.  Further, if a quadrupole is focusing (defocusing) in the $\hat{x}$-direction, it will be defocusing (focusing) in the $\hat{y}$-direction.  As a result, $\kappa_x = - \kappa_y$.  The strength of the quadrupole field is expressed as $\kappa_x = qG(z)(m \gamma \beta c)^-1$~\cite{wiedemann2015particle}. Here $q$ is the particle charge,
$G(z)$ is the spatially dependent quadrupole gradient in units of T/m, $\beta = v_z/c$, $v_z$ is the beam velocity, $c$ is the speed of light, and $\gamma = (1 - \beta^2)^{-1/2}$ is the relativistic factor. Both the quadrupole and dipole lattice forces are functions of distance $z$ along the reference trajectory. The quadrupole force profile is given by $G(z)$. The dipole-centripetal force is given by the profile function $1/R^2(z)$. The functional form of these profiles depends on the details of the dipole and quadrupole fields. The profile dependence is often treated approximately as a series of hard edge functions, and that is what we do here. A plot of the profiles of these two functions for 2 of the 36 sections (0.64~m) of UMER is shown in \cref{fig:kx_hard}. 

The focusing quadrupole field in the Warp simulation, but not the centroid model, includes small sextupole and octopole field corrections, whose whole multipole coefficients are $-0.1$ and $-0.05$, respectively. The quadrupole field structure induces linear forces in the transverse radial coordinate, while sextupole and octopole include quadratic and cubic terms, respectively. These corrections reflect experimental measurements of the first quadrupole in the 0.32~m lattice period, inducing small nonlinear forces\footnote{If these higher order moments are excluded, the error between Warp simulations with vs. without higher order moments grows linearly at about 10~{\textmu}m every turn, up to a value of 40~{\textmu}m after 5 turns.}.

\subsection{\label{sec:method:SINDy}Proposing Data-Driven Interpretable Models}

We prescribe a mathematical model based on the generic physics of accelerator lattices using periodic forcing obtained from simulation data. SINDy~\cite{de2020pysindy,Kaptanoglu2022} works by assuming one can model the evolution of some \hbox{$n$-dimensional} column vector $\mathbf{x}$ as a system of ordinary differential equations
\begin{equation} \label{eqn:feqn}
    \frac{d}{dt} \mathbf{x} = \mathbf{f}(\mathbf{x}).
\end{equation}
The variable $t$ in \cref{eqn:feqn} is the independent variable, $\mathbf{x}$ (we set $\mathbf{x} = [z,x_c,y_c]$ and detail why below) is the dependent variable, an observable either from simulation or experiment and $\mathbf{f}(\mathbf{x})$ is the $n$-dimensional equation governing how $\mathbf{x}$ evolves. We approximate $t\rightarrow z $ in our differential system by the paraxial approximation~\cite{reiser2008theory}.

We chose to use the directly observable variables $x_c$ and $y_c$ in SINDy. However, the $(x_c,y_c)$ parametric phase space plot is self-intersecting, \cref{fig:UMER_lat_phase}(b), and thus cannot be modeled by an autonomous, coupled system of differential equations, a fundamental assumption SINDy makes. We overcame this by adding the independent variable $z$ to the differential system as an additional dependent variable. With our $\mathbf{x} = [z,x_c,y_c]$ constituted, we now turn our attention to populating $\mathbf{f}(\mathbf{x})$.

Our proposed $\mathbf{f}(\mathbf{x})$ consists of three terms chosen solely from performing data analysis, which was agnostic toward the physical forcing function, and therefore the semi-analytic model detailed in Section~\ref{sec:centroid}. We did this to mimic an application environment where there is no detailed knowledge of the physical forcing function.

\begin{figure}[!htbp]
  \centering
  \includegraphics[width=\columnwidth]{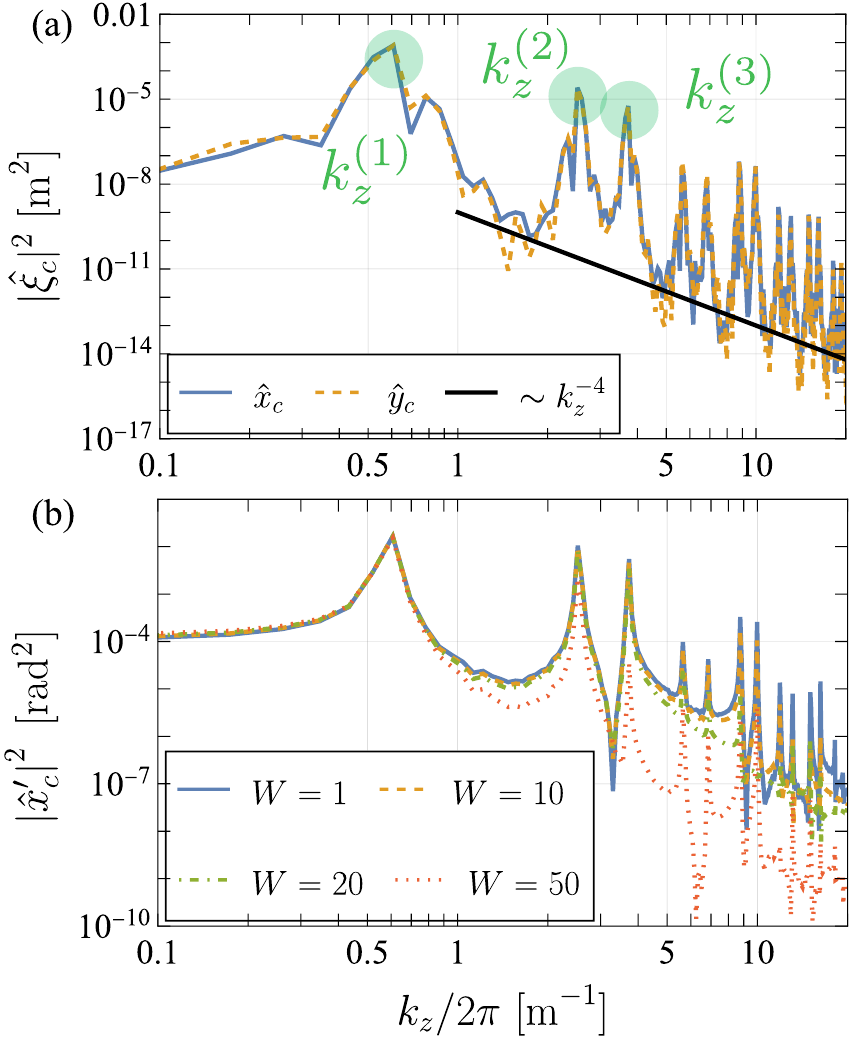}
\caption{\label{fig:Centroid_Fourier}(a) Log-log plot of the regularized power spectra magnitude for $|\hat{x}_c|$ solid blue and $|\hat{y}_c|$ dashed orange. The solid black line is a power law trend line for $k_z^{-4}$ created the numerical/finite length of the Fourier transform. \Cref{tab:kz_vals} details the value of the first three observed peaks for both $\hat{x}_c$ and $\hat{y}_c$ power spectra noted in transparent green circles. (b) Unregularized power spectra of the smoothed finite difference derivative of horizontal centroid motion $|\hat{x}_c|$ from \cref{fig:UMER_lat_phase}(b) for progressively larger uniform smoothing. The solid blue has a window size of 5~mm, dashed yellow 50~mm, dash-dot green 100~mm, and dotted red 250~mm.}
\end{figure}

The first term is a simple harmonic oscillator (SHO) based on \cref{fig:UMER_lat_phase}(b), which resembles SHO dynamics. The second is periodic forcing based on the first three peaks of the regularized power spectra of centroid dynamics and our knowledge that the lattice itself provides forcing, \cref{fig:Centroid_Fourier}(a). The third is a nonlinear interaction motivated by the behavior between the first three peaks, which is superimposed on the numerically induced power law in \cref{fig:Centroid_Fourier}(a). Appendix~\ref{sec:app:Fourier-data} details the regularization we used to produce the power spectra and enumerates features present within, \cref{fig:Centroid_Fourier}(a) guided by theoretical understanding. The individual $f_i$ are
\begin{align} 
    f_i = &\underbrace{\xi^{(0)}_{i} + \xi^{(1)}_{i,j} x_j}_{\text{SHO}}+ 
	\underbrace{\xi^{(c)}_{i,j} \text{cos}(k_j z) + \xi^{(s)}_{i,j} \text{sin}(k_j z)}_{\text{Periodic Forcing}} \nonumber + \\ &\underbrace{\xi^{(nc)}_{i,j m_{\text{max}} + j} x_j \text{cos}(k_m z)  +  \xi^{(ns)}_{i,j m_{\text{max}} + j} x_j \text{sin}(k_m z)}_{\text{Nonlinear Interaction}} \, , \label{eqn:fi}
\end{align}
where $i \in (0,1,2)$, $j\in(0,1,2)$, $m\in(0,1,2)$, and $m_{\text{max}}=3$. The quantity  $\xi_{a,b}^{(...)}$ is an indexed coefficient with four entries for the SHO terms, six for the periodic forcing, and eighteen from the nonlinear interaction totaling 28, $n_b$, basis functions, and $k_j$ is the $j$'th injected peak from the data's power spectra, \cref{fig:Centroid_Fourier}(a) and \cref{tab:kz_vals}. The parenthetical superscript for each $\xi$ coefficient denotes which term those coefficients are associated with.  \Cref{tab:kz_vals} details the three lowest order modes of the data's power spectra, which we inject into our basis function matrix $\Theta(\mathbf{X})$.

We measure $\mathbf{x}$ at $m$ equidistant times and place the measured values of $\mathbf{x}$ into a matrix $\mathbf{X}$:
\begin{eqnarray*}
    \mathbf{X} = \begin{bmatrix} 
    \mathbf{x}^T(t_1) \\
    \vdots  \\
    \mathbf{x}^T(t_m)
    \end{bmatrix}
    = \begin{bmatrix}
    x_1(t_1)  & \dots & x_n(t_1)  \\
    \vdots & \ddots & \vdots \\
    x_1(t_m) & \dots & x_n(t_m)
    \end{bmatrix} \; .
\end{eqnarray*}
We differentiate the matrix, $d\mathbf{X}/dt = \dot{\mathbf{X}}$ which we use in SINDy. We create a matrix $\Theta(\mathbf{X})$ made of chosen basis function, \cref{eqn:fi} with $m$ rows and $n_b$ columns. The matrix $\dot{\mathbf{X}}$ is equated to $\Theta(\mathbf{X})$ time a coefficient matrix $\boldsymbol{\Xi}$ with $n_b$ rows and $n$ columns which we solve given an optimization technique
\begin{eqnarray*}
\dot{\mathbf{X}} 
    = \Theta(\mathbf{X}) \boldsymbol{\Xi} \, ,
\end{eqnarray*}
where the matrix $\Theta(\mathbf{X})$ is 
\begin{eqnarray*}
\Theta(\mathbf{X})
    = \begin{bmatrix}
   \mathbf{1} & \mathbf{X} & \text{cos}(\mathbf{k} z) &  \text{sin}(\mathbf{k} z) & \mathbf{X} \text{cos}(\mathbf{k} z) & \mathbf{X} \text{sin}(\mathbf{k} z)
    \end{bmatrix} \, .
\end{eqnarray*}
with $\mathbf{1}$ a column vector of ones of length $m$, and $\mathbf{k}$ the three-dimensional column vector of injected lowest order peaks from the power spectra, \cref{fig:Centroid_Fourier}(a) and \cref{tab:kz_vals}. \Cref{tab:kz_vals} details the three lowest order lattice modes of the data which are injected into our basis function library matrix $\Theta(\mathbf{X})$.

The $\boldsymbol{\Xi}$ matrix has 84 entries, which equals the number of proposed basis function $n_b$ (28) times the number of state variables $n$ (3)
\begin{equation*}
    \boldsymbol{\Xi}^T = \begin{bmatrix}
\boldsymbol\xi^{(0)} & \boldsymbol\xi^{(1)}& \boldsymbol\xi^{(c)}& \boldsymbol\xi^{(s)} & \boldsymbol\xi^{(nc)} & \boldsymbol\xi^{(ns)} 
\end{bmatrix} \, .
\end{equation*}
Values for $\boldsymbol{\Xi}$ are shown in Appendix~\ref{sec:app_Coef_Tab}.

\begin{table}[b]
	\setlength\tabcolsep{3.5pt}
	\centering
	\caption{\label{tab:kz_vals}Three lowest order power spectra peaks  centroid dynamics from \cref{fig:Centroid_Fourier}(a). The parenthetical superscript is the index $i$. The quantities $k_z^{(i)}$ and $L_z^{(i)}$ are the angular frequency(wavenumber) and corresponding wavelength. With knowledge of the underlying lattice structure, we note that $L_z^{(1)}$ is the return time for the harmonic oscillator and that the average of $L_z^{(2)}$ and $L_z^{(3)}$ is UMER's FODO period. This is discussed in more detail in Appendix~\ref{sec:app:Fourier-data}.}
 \begin{ruledtabular}
	\begin{tabular}{lcr}
		$i$ & $k_{z}^{(i)}/2\pi$ [1/m] & $L_{z}^{(i)}$ [m]\\ \hline
		1 & 0.61 & 1.64 \\ 
		2 & 2.52 & 0.397 \\
		3 & 3.62 & 0.268 \\
	\end{tabular}
  \end{ruledtabular}
\end{table}

The simulation data we used to produce $\dot{\mathbf{X}}$ includes hard-edge effects from the quadrupoles and dipoles. To avoid spurious results in SINDy's fitting process, we smoothed the data with a uniform filter of varying window size $W \in (1,10,20,50)$ denoting $dz$ consecutive measurements corresponding to 5~mm, 50~mm, 100~mm, and 250~mm,  respectively. We chose a window size of $W=10$ that was sufficiently smooth for $x_c'$ yet retained high fidelity to the proposed FOM, \cref{fig:Centroid_Fourier}(b).

\section{\label{sec:res}SINDy Results}

We obtained the results described in this section by using SINDy's Sequentially Threshold Least Squares (STLSQ)~\cite{brunton2016discovering} optimization algorithm. We have chosen SINDy's qualitative reproduction of the data's power spectra as a Figure of Merit (FOM), giving indication of whether the fit captures the data's dynamics. We detail three fits with progressively more basis functions and report their capturing of the FOM of the training data over one turn. We probe the optimization hyperparameter stability of the fit with the most basis functions and highest fidelity.  We assess the proposed model's prediction robustness and dynamic fidelity over two additional turns with three different hyperparameter cases.

\begin{figure}[!htbp]
  \centering
  \includegraphics*[width=\columnwidth]{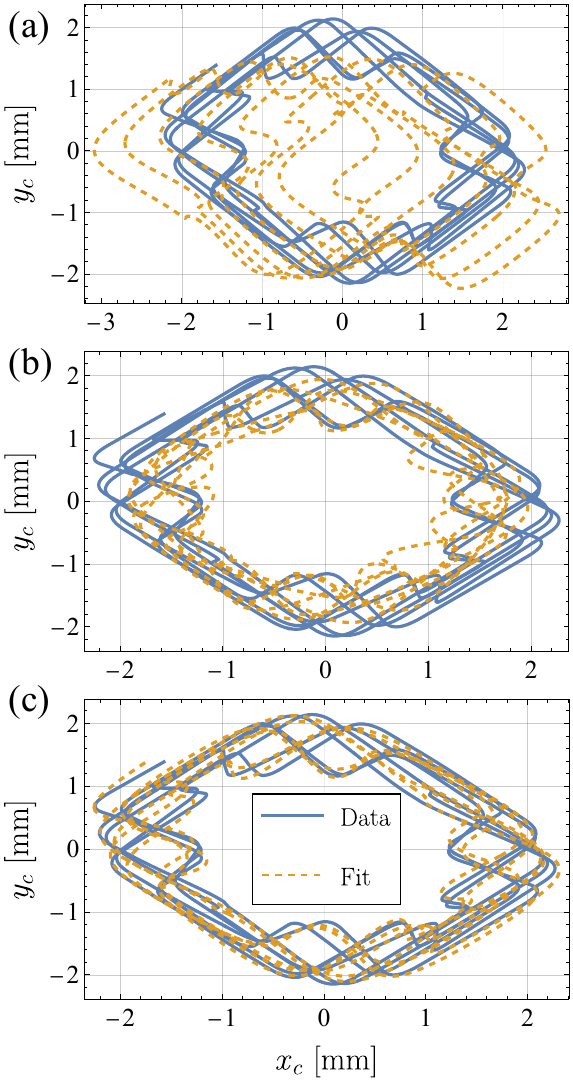}
\caption{\label{fig:SINDy_param}Parametric phase space plots with the solid blue line simulation data and the dashed yellow the SINDy fit. (a) fit 1 from Sec.~\ref{sec:res:fit1}, (b) fit 2 from Sec.~\ref{sec:res:fit2}, and (c) fit 3 from Sec.~\ref{sec:res:fit3}.}
\end{figure} 

\subsection{\label{sec:res:fit1}Fit 1 - Periodic Forcing}

This case contains only the injected first three modes, \cref{tab:kz_vals}, from the simulation data's power spectrum, \cref{fig:Centroid_Fourier}(a)
\begin{equation} 
    f_i =  \xi^{(c)}_{i,j} \text{cos}(k_j z) + \xi^{(s)}_{i,j} \text{sin}(k_j z)  \label{eqn:fit1}
\end{equation}
with tabulated values for $\xi_{i,j}^{(c)}$ and $\xi_{i,j}^{(s)}$ in \cref{eqn:fit1} reported in Appendix~\ref{sec:app_Coef_Tab}'s \cref{tab:model_coefs_xc,tab:model_coefs_yc}. 

\Cref{fig:SINDy_param}(a) shows a comparison between the simulation data and the SINDy fit using only the periodic forcing terms, with favorable agreement for the $y_c$ coordinate and poorer agreement for $x_c$. The power spectra, \cref{fig:SINDy_fourier}(a), capture injected mode amplitude, yet errors remain. Specifically, the constant term near the axis of symmetry, the valleys between the injected modes are off from the data's value by orders of magnitude, and the twin peak pair cascade existing at linear frequencies exceeding 5~m$^{-1}$. However, we only included terms with linear frequencies on the inverse length scale of 3.7~m$^{-1}$, so we cannot expect the model to capture higher frequencies. Inspecting the centroid model in Section~\ref{sec:centroid} we note that the periodic forcing terms are inadequate to capture the phase space dependence included by the $\xi_c = \pm \kappa_{\xi} \xi_c$ centroid equations. Further dependence on phase space location ($x_c$ or $y_c$) is needed.

\subsection{\label{sec:res:fit2}Fit 2 - SHO and Periodic Forcing}

This fit iteration contains both the periodic forcing terms from \cref{tab:kz_vals} and the SHO terms
\begin{equation} 
    f_i = \xi^{(0)}_{i} + \xi^{(1)}_{i,j} x_j + 
	\xi^{(c)}_{i,j} \text{cos}(k_j z) + \xi^{(s)}_{i,j} \text{sin}(k_j z)  \, . \label{eqn:fit2}
\end{equation}
with tabulated values for $\xi_{i,j}^{(c)}$, $\xi_{i,j}^{(s)}$, $\xi_{i}^{(0)}$, and $\xi_{i}^{(1)}$ in \cref{eqn:fit2} reported in Appendix~\ref{sec:app_Coef_Tab}'s \cref{tab:model_coefs_xc,tab:model_coefs_yc}. 

\Cref{fig:SINDy_param}(b) shows a comparison between the simulation data and the SINDy fit using the periodic forcing and SHO terms. We obtain better agreement for the $x_c$ and $y_c$ coordinates over the results obtained in Section~\ref{sec:res:fit1}. In contrast to prior results, this fit's dynamics are entirely within the training data, and do not wander like the preceding fit, Sec.~\ref{sec:res:fit1}. Our inclusion of SHO terms enabled SINDy to find a more constrained model for $x_c'$ and $y_c'$. The coupling between $x_c'=y_c$ and $y_c' = -x_c$, see \cref{tab:model_coefs_xc,tab:model_coefs_yc} in Appendix~\ref{sec:app_Coef_Tab}, simulates a harmonic oscillator as achieved in the uncoupled centroid model in Section~\ref{sec:centroid} from the second order differential equations.

The power spectra shown in \cref{fig:SINDy_fourier}(b) captures injected mode amplitudes and the constant term as $k_z/2\pi \rightarrow 0$. Still, the error between the injected periodic forcing is present, in addition to the lack of capturing at linear frequencies exceeding 5~m$^{-1}$. Terms including nonlinear interaction between the periodic forcing are yet to be included in the model; the next, and highest fidelity, fit detailed in Section~\ref{sec:res:fit3} includes nonlinear interaction into $f_i$.

The natural angular frequency $k_z$ of the SHO can be found by multiplying the $y_c$ coefficient in $d x_c /dz$ and the $x_c$ coefficient in $d y_c /dz$, \cref{tab:model_coefs_xc,tab:model_coefs_yc}. The corresponding wavelength to this natural angular frequency is the return time for individual centroid coordinates, $\sim 1.6$~m, \cref{fig:UMER_lat_phase}(b). This result reinforces the fact that SINDy uncovered part of the underlying dynamics.  

\subsection{\label{sec:res:fit3}Fit 3 - SHO, Periodic Forcing, and Nonlinear Interaction}

This section details the results with the full prescribed $f_i$ equation detailed in Section~\ref{sec:method:SINDy}. \Cref{eqn:fi} is reproduced below for the reader's convenience
\begin{align*} 
    f_i = &\xi^{(0)}_{i} + \xi^{(1)}_{i,j} x_j + 
	\xi^{(c)}_{i,j} \text{cos}(k_j z) + \xi^{(s)}_{i,j} \text{sin}(k_j z) + \\ &\xi^{(nc)}_{i,j m_{\text{max}} + j} x_j \text{cos}(k_m z)  +  \xi^{(ns)}_{i,j m_{\text{max}} + j} x_j \text{sin}(k_m z)  \, .
\end{align*}
The tabulated values for $\xi_{i,j}^{(...)}$ are reported in Appendix~\ref{sec:app_Coef_Tab}'s \cref{tab:model_coefs_xc,tab:model_coefs_yc} for Fit 3.

\begin{figure}[!htbp]
  \centering
  \includegraphics*[width=\columnwidth]{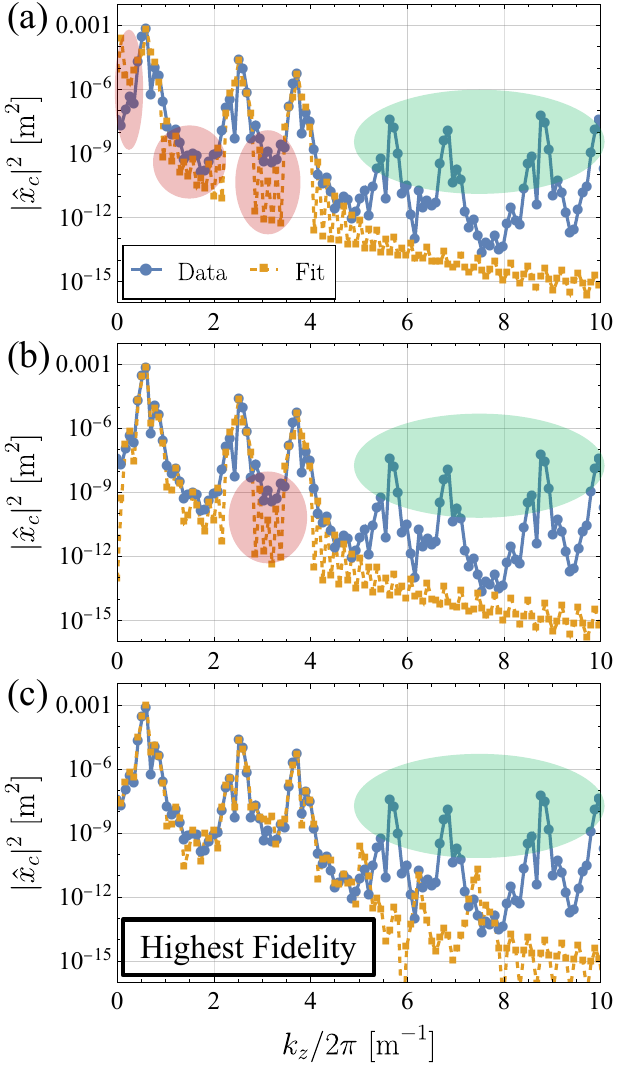}
\caption{\label{fig:SINDy_fourier}Semi-log plot of the $\hat{x}_c$ power spectra for the simulation data in the solid blue line with circles and the SINDy fit in the dashed yellow with squares. (a) fit 1 from Sec.~\ref{sec:res:fit1}, (b) fit 2 from Sec.~\ref{sec:res:fit2}, and (c) the highest fidelity result fit 3 from Sec.~\ref{sec:res:fit3}. The red circles denote progressively eliminated deficiencies in the SINDy fits on the injected length scales. The green circles show the peak pair cascade that is not captured in the SINDy fits.}
\end{figure}

This proposed fit evolution obtained the highest fidelity using SINDy. Phase space dynamics show good agreement between the Warp data and the SINDy fit \cref{fig:SINDy_param}(c). The FOM is well captured \cref{fig:SINDy_fourier}(c), with each peak's amplitude and behavior on either side of that peak captured as well. However, the cascading pairs of peaks observed in \cref{fig:SINDy_fourier}(c) are conspicuously absent. This disagreement is not replicated in the centroid model. The reasons for this are discussed later, and methodologies one could use to allow greater capturing of statistics are discussed in Sec.~\ref{sec:disc} and App.~\ref{sec:app:Fourier-data}.

\subsection{\label{sec:res:pred}Prediction Ability Case Study}

\begin{figure}[!htbp]
  \centering
  \includegraphics*[width=\columnwidth]{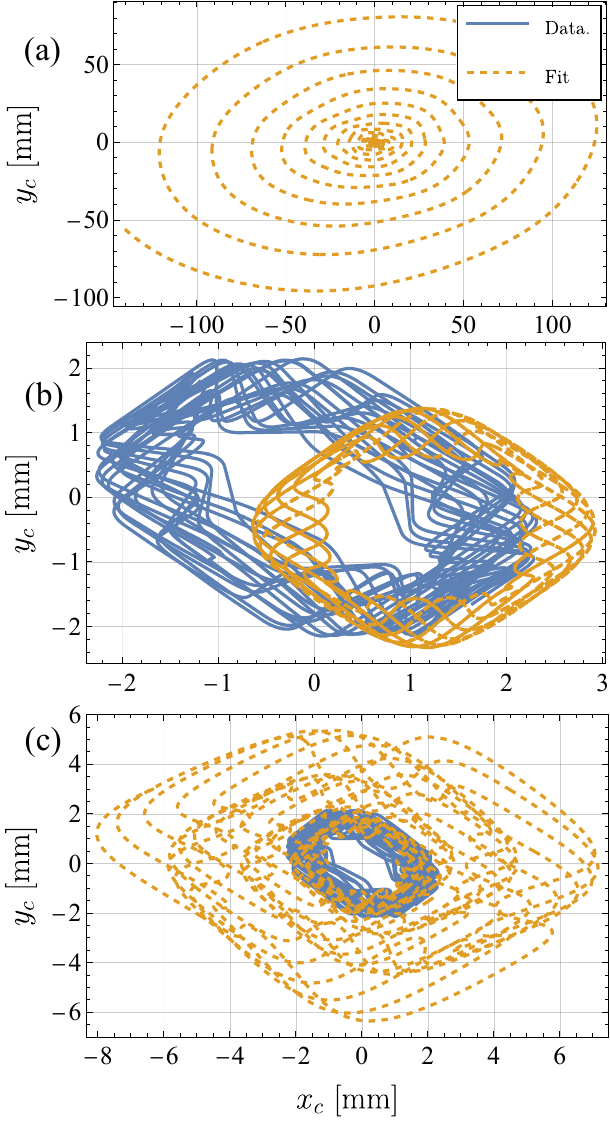}
\caption{\label{fig:case_comp}Different optimization results (a-c) plotted in parametric phase space over three lattice turns. (a) Case 1 with a threshold=0.0 and $\alpha=0.1$, (b) case 2 with a threshold=1.0e-3 and $\alpha=0.25$, and (c) case 3 with a threshold=1.0e-4 and $\alpha=0.05$.}
\end{figure}

In this section, we examine SINDy's prediction ability using the proposed evolution equation for fit 3. We performed a systematic logarithmic search (points taken at different order magnitudes) in the threshold value and a linear search in the regularization parameter but noticed no patterns in that search. Also, the fitting results proved sensitive to optimization hyperparameters. As such, we only report three prediction cases, see \cref{fig:case_comp}, over two additional turns in the UMER lattice. Each case employed different optimization hyperparameters to show the evolution equations' hyperparameter sensitivity. Case 1 has the same optimization hyperparameters as fit 3, while case 2 and case 3 have different hyperparameters. The STLSQ algorithm SINDy employs has two hyperparameters. These are a lower threshold value for learned coefficients of the evolution equations and a regularization parameter $\alpha$ that constraints disparate coefficient values. Other optimization schemes other than STLSQ were not used to limit optimization hyperparameter dimensionality. 

\begin{figure}[!htbp]
  \centering
  \includegraphics*[width=\columnwidth]{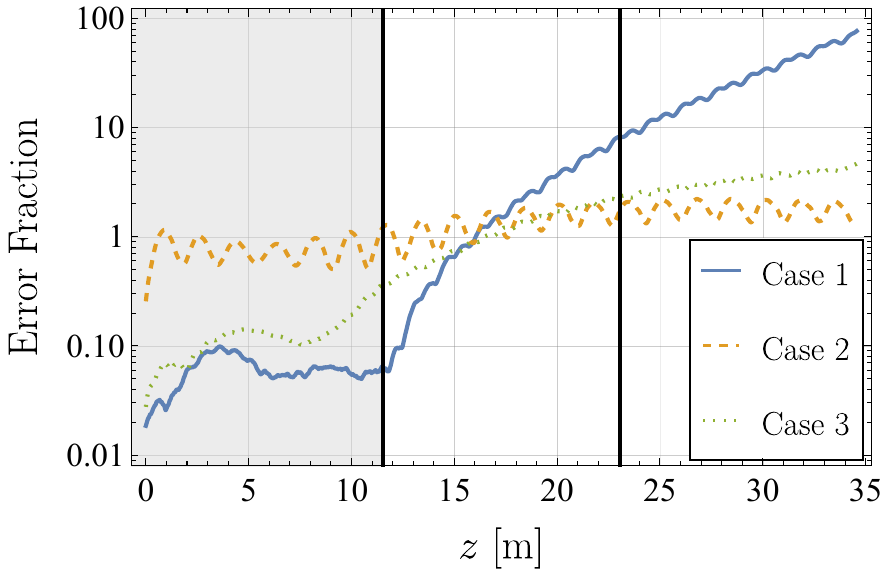}
\caption{\label{fig:case_error}The error fraction of \cref{fig:case_comp} plotted with semilog axis plotted over three turns of the lattice. The black, vertical lines mark sequential turns. The shaded grey area is the training set; the further two turns are predictions. The solid blue line is case 1, dashed yellow case 2, and dash-dot green case 3.}
\end{figure} 

\Cref{fig:case_comp}(a) shows the results of fit 3 (noted as case 1) integrated for two further turns. This fit is unstable, which comes the positive $y_c$ term in the evolution equation, \cref{eqn:fi} reported in \cref{tab:model_coefs_yc}. This linear term introduces exponential growth of the $y_c$ coordinate, which bleeds over into $x_c$ and causes the whole solution to grow without bound. The hyperparameters used for this case had a regularization parameter $\alpha=0.1$ and a threshold of 0.0, allowing all terms in the $\boldsymbol{\Xi}$ matrix to be nonzero. This represents overtraining on the data for prediction purposes. While the obtained FOM from \cref{fig:SINDy_fourier}(c) is recovered almost exactly for training, prediction is unstable.

Conjecturing that overtraining (overfitting) on the training data causes solution instability, we perform another optimization while restricting the terms of $\boldsymbol{\Xi}$. We increased the threshold to 1e-3 and the regularization parameter to 0.25, \cref{fig:case_comp}(b). This result has well-contained phase space behavior with a systemic offset induced by only having a small, realizable subset of terms in the governing evolution equation \cref{eqn:fi}, which are three of linear cosine and sine periodic forcing functions.

Aiming to balance the two extremes between case 1 and case 2, we decreased the threshold to a smaller, though still nonzero, value of 1e-4. To further allow disparate coefficient sizes---motivated by the nonlinear interaction from \cref{fig:Centroid_Fourier}(a)---we set the regularization parameter $\alpha$ to 0.05, \cref{fig:case_comp}(c). While the unbounded growth observed in Case 1's \cref{fig:case_comp}(a) isn't present, the solution can't track dynamics over a longer period.

We quantify the error between the fit and data with an error fraction 
\begin{equation*}
    \text{Error Fraction} = \sqrt{\frac{   \left( x_{c,\text{S}} - x_{c,\text{D}} \right)^2 + \left( y_{c,\text{S}} - y_{c,\text{D}} \right)^2     } {\sigma^2_{x_{c,\text{D}}} + \sigma^2_{y_{c,\text{D}}}}  } \, .
\end{equation*}
The error fraction is the square root of the sum of the square of the errors (SINDy minus simulation data) divided by the sum of the squares of the standard deviation $\sigma_{\xi_{c,\text{D}}}$ of the simulation. The subscript S(D) corresponds to SINDy(data). \Cref{fig:case_error} reports the results of the error fraction over three turns. The bounded error fraction of case 2 indicates long-term stability.

With the goal of long term solution stability and prediction in mind, we introduce an additional statistical FOM, the dimensionless ratio of the difference in the second moments of the prediction to the actual moments of the beam centroid evolution. The first moments were not considered, as the data's first moments are near zero, causing relative differences to amplify misleadingly. This FOM is evaluated over the three turns shown in \cref{fig:case_comp}. \Cref{tab:metrics} reports the results for all these values. Note again that case 2 obtains favorable long term results over successive turns.

\begin{table}[!htbp]
	\setlength\tabcolsep{3.5pt}
	\centering
	\caption{\label{tab:metrics}Dimensionless second moment ratios for various turns predicted by SINDy compared to Warp simulation data. Successive left-to-right entries in the third and fourth columns denote the dimensionless ratio evaluated from the beginning of turn 1 to the end of turn 1, 2, or 3, respectively.}
 \begin{ruledtabular}
	\begin{tabular}{lccr}
	      Case & (threshold,$\alpha)$ &  $\dfrac{\Delta \langle x_c^2 \rangle}{\langle x_c^2 \rangle}$ & $\dfrac{\Delta \langle y_c^2 \rangle}{\langle y_c^2 \rangle}$\\ \hline
		1 & (0.0, 0.1) & 0.05, 9.83, 568.0 & 0.0239, 4.11, 292 \\ 
		2 & (1e-3,0.25) & 0.340, 0.346, 0.348 & 0.176, 0.194, 0.190  \\
		3 & (1e-4,0.05) & 0.106, 1.06, 3.42 & 0.080, 0.650, 2.27  \\
	\end{tabular}
  \end{ruledtabular}
\end{table}

Case 2 is the most robust prediction fit per the error fraction, \cref{fig:case_error}, and dimensionless second moment metrics, \cref{tab:metrics}. This fit only has terms that won't grow without bound in the $(x_c, y_c)$ phase space. In contrast, case 1 has terms that grow without bound in the prediction phase and case 3 is not able to predict long-term dynamics within the constrained area realized by simulation data. Turning to reality, we integrate case 2 forward for 100 total turns to evaluate long-term prediction stability, a realizable number of turns in UMER.

\begin{figure}[!htbp]
  \centering
  \includegraphics*[width=\columnwidth]{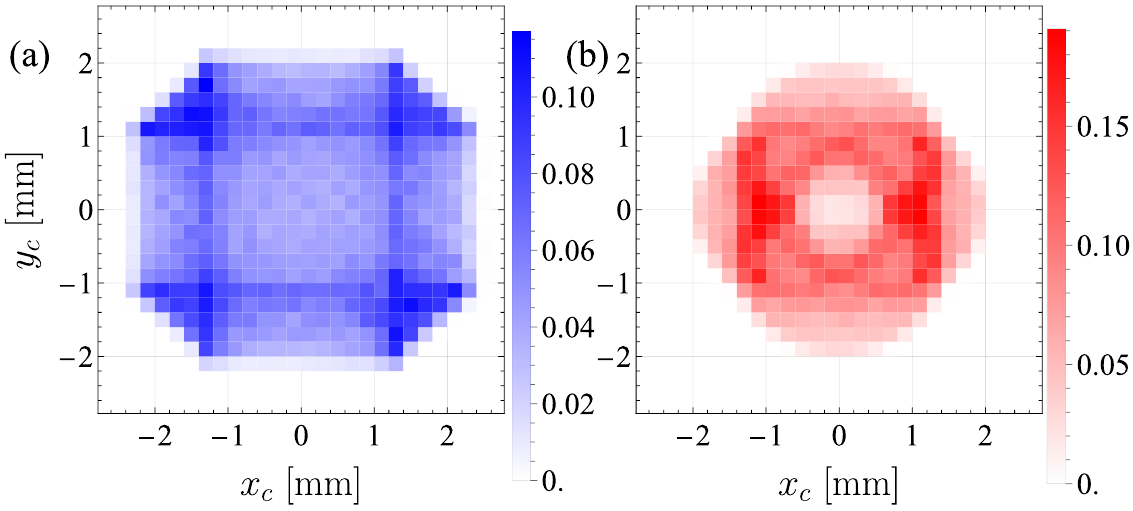}
\caption{\label{fig:phase_100_turn}Two-dimensional $(x_c,y_c)$ histograms for probability whose color bars represent probability per 200~{\textmu}m$^{2}$ bin size over the range $x_c,y_c \in (-2.5,2.5)$~mm. (a) The simulation data taken over 100 turns. (b) SINDy fit taken over 100 turns.  Note how the fit data is contained within the area visited by the Warp simulated beam.}
\end{figure} 

\Cref{fig:phase_100_turn}(a) reproduces the SINDy fit found in Sec.~\ref{sec:res:fit2} and replots the data onto a binary two-dimensional histogram showing trajectory presence. Note how the ensemble area of the fit is contained with the training data. \Cref{fig:phase_100_turn}(b) shows the solution from Sec.~\ref{sec:res:fit2} integrated forward for 99 additional prediction turns, 100 turns in total, and compared to simulation data for 100 turns. Note how the fit is well-bounded within the ensemble area of the training data; we have obtained robust prediction for a particular hyperparameter set with a minimal physical description. This demonstrates SINDy's ability to discover underlying beam dynamics using an intuitive, data-driven approach necessitated for realistic applications where no analytic solutions exist.

\section{\label{sec:centroid}Semi-Analytic Centroid Dynamics}

A motivating reason that we chose to apply SINDy to a misinjected beam in UMER is so results can be verified against a centroid motion model. Recall we prescribed SINDy's evolution terms without foreknowledge of the centroid model, and as such cannot expect the SINDy fits to outperform the model which has first-principle physics. In addition, the Warp simulations SINDy trains on have both numerical and statistical errors arising from discretization and a finite number of simulation macro-particles, respectively. 

\begin{figure}[!htbp]
  \centering
  \includegraphics*[width=\columnwidth]{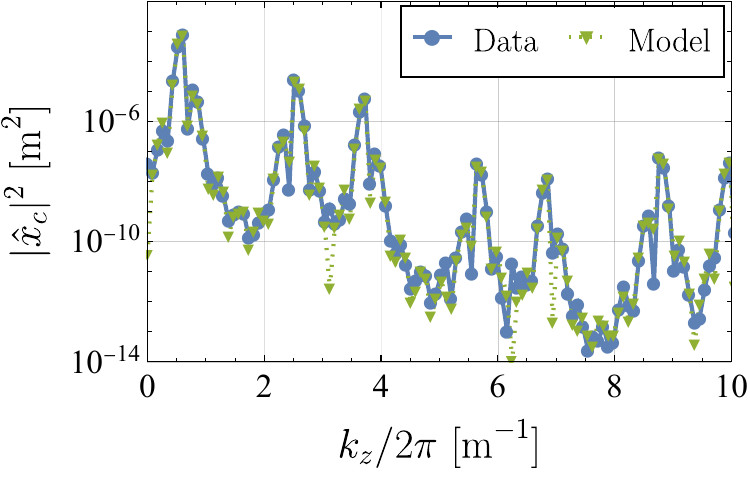}
\caption{\label{fig:xc_yc_mod_fourier}Power spectra $| \hat{x}_c |^2$ for Warp simulation data in solid blue for $\hat{x}_c$ with solid circles and the centroid model in dotted green for $\hat{x}_c$ with down-pointing triangles.}
\end{figure}

\Cref{fig:xc_yc_mod_fourier} shows the power spectra of \cref{eqn:xc_cent,eqn:yc_cent} solved with initial conditions of $x_c(0)=y_c(0)=$ 1~mm and $x_c'(0)=y_c'(0)=$ 0~mrad. The forcing function for the semi-analytic model were the same as was used for the simulation Warp up to linear order, and are shown for the first two periods in \cref{fig:kx_hard}. Recall that the simulation in Warp included small sextupole (quadratic) and octopole (cubic) corrections for the first quadrupole of every 0.32~m period reflecting experimental measurements. 

We observe the centroid model solution replicating beam dynamics from Warp in both the parametric phase space and power spectra, \cref{fig:xc_yc_mod_fourier}, over one turn. As expected, the centroid model captures the peak pair cascade as observed in the simulation data, since the direct lattice physical forcing functions were used in the model.

A key difference between the centroid model and the proposed SINDy fits detailed proposed in Section~\ref{sec:method:SINDy} and detailed in Sections~\ref{sec:res:fit1}-~\ref{sec:res:fit3} is the order of differential equations. The centroid model detailed in this section has the second-order differential equations for $x_c$ and $y_c$, \cref{eqn:xc_cent} and \cref{eqn:yc_cent}, decoupled from each other, while the proposed SINDy fit enforces coupling between $x_c$ and $y_c$ via the SHO term, \cref{eqn:fi} and are first order differential equations. The SINDy fit replicates the second-order nature of centroid motion by coupling the two transverse centroid coordinates together. Another key difference between the centroid model and the SINDy fits is that the model has directly injected into its evolution equations the real physical forcing functions. Because of these differences, we cannot expect the SINDy model to outperform the centroid model for either qualitative or statistical recapturing of beam dynamics.  

\section{\label{sec:allcomp}Comparisons Between Centroid Model, SINDy Fit, and Simulation Data}

\begin{figure}[!htbp]
  \centering
  \includegraphics*[width=\columnwidth]{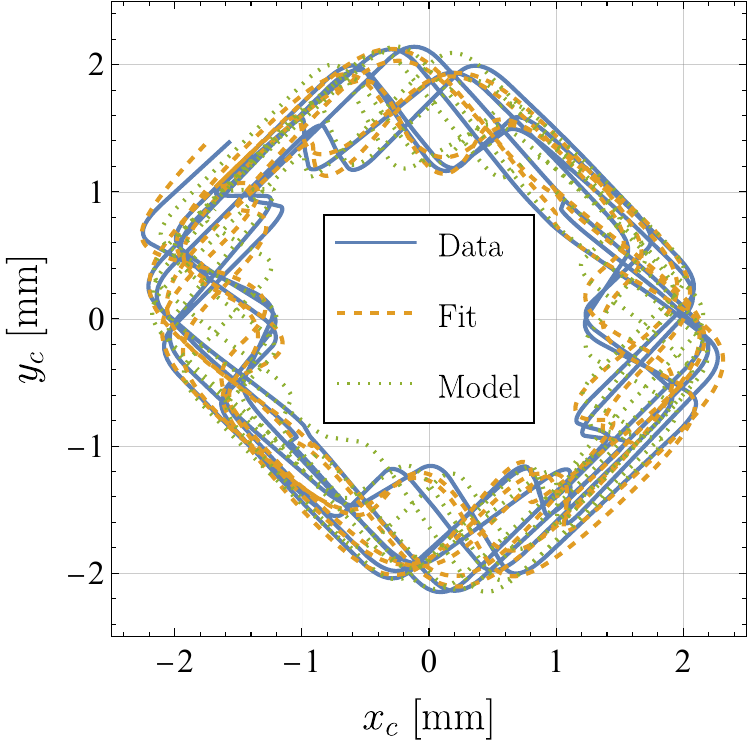}
\caption{\label{fig:phase_all_comp_turn1}Parametric plot $(x_c,y_c)$ comparison for one turn between Warp simulation data (solid blue), SINDy fit 3/case 1 from Sec.~\ref{sec:res:fit3} (dashed yellow), and the centroid model (green dash-dot).}
\end{figure}

\Cref{fig:phase_all_comp_turn1} shows the solutions obtained from each method plotted over one turn (fit 3 SINDy). Overall, both the fit and the model capture the dynamics of the $x_c$ and $y_c$ centroid coordinates\footnote{It is interesting to note that on the timescale of 5 turns the quantity $x_c \times y_c$ becomes negative on average, and this is what both the fit and the model capture. On longer time scales, this quantity precesses about the center of the beam pipe on a length scale of twenty turns ($\sim$200~m). All SINDy fits shown thus far do not capture this secular behavior, whereas the centroid model unfailingly does so, to the longest simulations we have run on the order of 100 turns or $\sim1$~km.}, an indication that both capture appropriate physics.

\begin{figure*}[!htbp]
  \centering
  \includegraphics*[width=\textwidth]{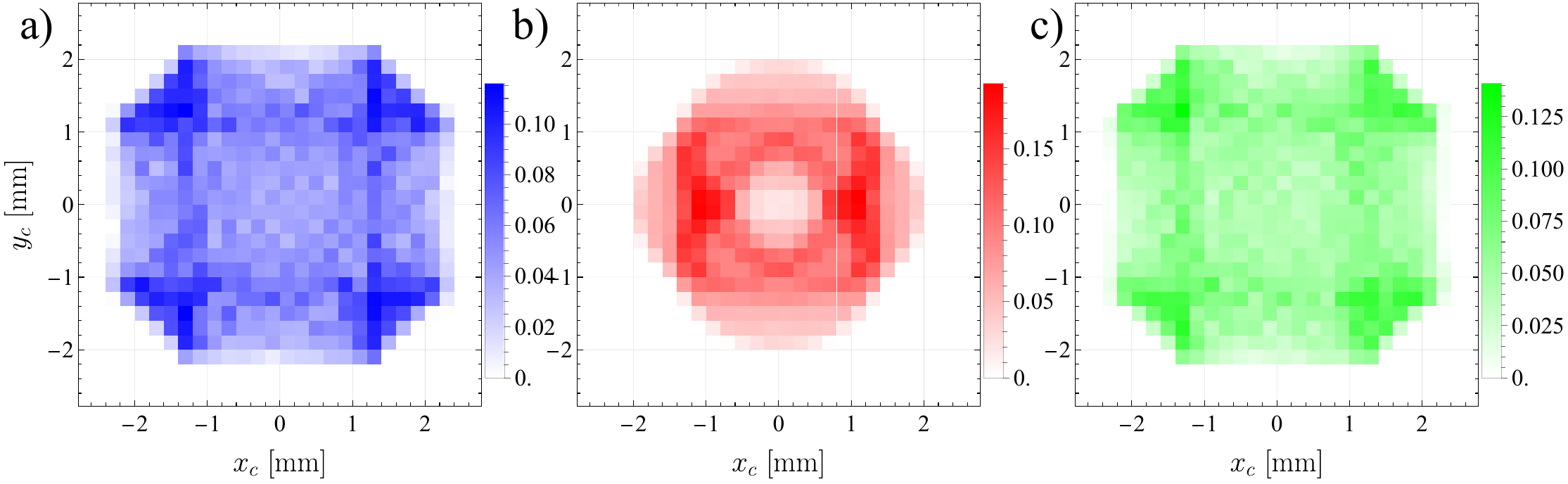}
\caption{\label{fig:phase_all_comp}Two-dimensional phase space probability histogram whose color bars represent probability per 200~{\textmu}m$^{2}$ bin size for a testing set over $\sim85$ lattice turns. \Cref{fig:phase_100_turn}(a-b) are reproduced here for the reader's convenience. (a) Simulation data, (b) the SINDy fit 2 from Sec.~\ref{sec:res:fit2}, and (c) the centroid model deployed in Sec.~\ref{sec:centroid}. Note the correlation in the histogram probability amplitude and structure between (a) and (c).}
\end{figure*}

The SINDy fit as noted previously compares favorably with the data over one turn, \cref{fig:phase_all_comp_turn1}, while the centroid model replicates the data fairly well over one turn, though not as exactly as this particular SINDy model. However, when integrated over 85 turns, the centroid model outperforms the SINDy model, capturing the phase space ensemble mapped out in $(x_c, y_c)$, \cref{fig:phase_all_comp}(a-c). The centroid model, \cref{fig:phase_all_comp}(c), captures the areas around the points $(\pm1,\pm1)$ where the centroid is likely to reside is captured well in the centroid model. The SINDy fit, \cref{fig:phase_all_comp}(b), is robust, as evidenced by its residence in the ensemble phase space areas well contained within the simulation data's ensemble data. 

\begin{figure}[!htbp]
  \centering
  \includegraphics*[width=\columnwidth]{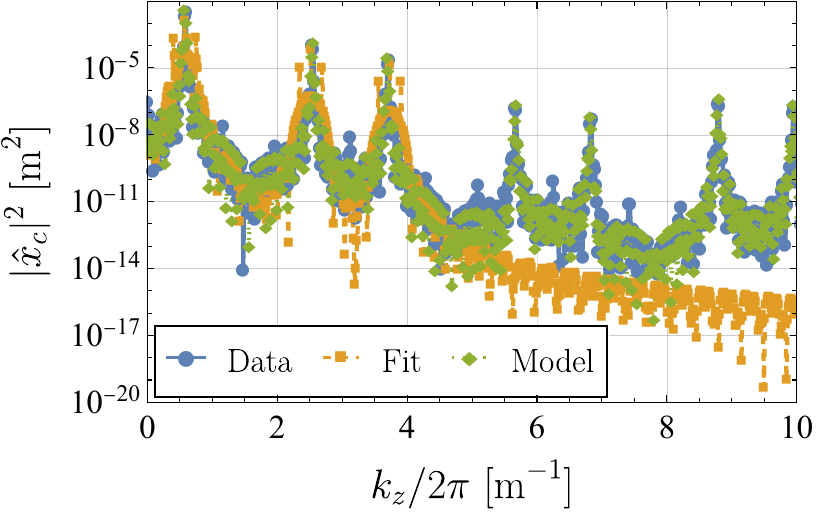}
\caption{\label{fig:fourier_all}Power spectra for $\hat{x}_c$ over $k_z/2\pi \in (0.0,10.0)$ of simulation data in solid blue with circles, SINDy fit in dashed yellow with squares, and the centroid model dotted green with diamonds and open circles.}
\end{figure}

Statistically, the centroid model outperforms SINDy by reproducing the power spectra of $x_c$ and $y_c$, \cref{fig:fourier_all} at higher frequencies. This is to be expected, though, as we did not inject frequencies of higher than 3.7~m$^{-1}$ into the SINDy evolution equations. The SINDy fit reproduces the magnitude and structure between the lowest order periodic forcing as those modes were directly injected into the model. The centroid model captures the cascading pairs of peaks, which are readily visible in the semi-log plots \cref{fig:fourier_all}. We note that these pairs of peaks are not individual harmonics, each pair is a harmonic of the interaction of the natural linear frequency of the system $k_{N}/2\pi$, the SHO lap time of 0.6~m$^{-1}$, and harmonics of the forcing linear frequency $n \, k_{F}/2\pi$ of the FODO lattice 3.12~m$^{-1}$ inverse period, where $n$ is the order of the harmonic. The interaction of subsequent pairings, $( n \, k_F \pm k_N )/ 2 \pi$, of these quantities is what we see as cascading peak pairs in the power spectra. This is discussed in further detail in App.~\ref{sec:app:Fourier-data}.

\section{\label{sec:disc}Discussion} 

Within this paper, we have examined SINDy's effectiveness in reproducing beam dynamics. We accomplished this using a data-driven model that was created with terms motivated by intuition and data analysis, without recourse to an underlying dynamical model, nor with explicit reference to the periodic focusing lattice structure, so foreknowledge of explicit terms would not infiltrate our prescribed basis functions. We chose to implement SINDy with these constraints to mimic an application environment where analytic models don't exist or where lattice elements are unknown or cannot be modeled. Because we chose a well-established lattice with a simple beam, we were able to employ a semi-analytic centroid model and interpret SINDy's fit results a posteriori. 

The SINDy fits exhibited higher fidelity to the training data as we included progressively more terms, yet became less resilient to different optimization hyperparameters during prediction, Sec~\ref{sec:res:pred}. This outcome could come from: 
\begin{itemize}
    \item the training data supplied to SINDy being statistically non-stationary;
    \item the amount of the data presented during training, which consisted solely of 6000 data points (three variables $\mathbf{x}=[z,x_c,y_c]$ over an 11.52~m circumference UMER lattice measured every 5~mm);
    \item the prescribed basis functions (specifically the periodic forcing, \cref{eqn:fi}, which were chosen before each particular peak was identified using insight from the centroid model);
    \item and the dimensionality of SINDy's differential system (two-dimensional vs the four dimensional, though decoupled, centroid model). 
\end{itemize} 
Nevertheless, even with these synthetically imposed impediments, SINDy captured the physics of both: the natural frequency of the system $k_N$ by coupling the coordinates $x_c$ and $y_c$ together, artificially reproducing the appropriate two-dimensional SHO motion arising from the inherent misinjected beam dynamics; and the directly injected periodic focusing modes using $k_2$ and $k_3$ from \cref{tab:kz_vals}. However, SINDy was unable to capture the dynamics of the peak pair cascade. This, although disappointing, is not surprising in the light that we only three (and of inappropriate frequencies, nonetheless) periodic forcing terms coupled with $x_c$ and $y_c$ up to the linear frequency 3.7~m$^{-1}$. We remark that SINDy was able to project the nonlinear dynamics present onto (admittedly inaccurate) basis functions capturing the underlying dynamics. Basis function refinement is a subject of future work.

A primary reason we chose to use the first turn for training was to compare to a similar, though not interpretable, prediction technique from the field of ML~\cite{komkov2019reservoir}. While these ML results obtained a favorable comparison to the simulation data, the dimensionality of their differential system was four thousand, versus SINDy's two, and the centroid model's four. However, we note their results outperformed SINDy for prediction over those turns for the cases we have shown within this paper, \cref{fig:case_comp}. Though, for different settings of the STLSQ optimization, SINDy produced differential systems comparable in prediction capability to the ML technique~\cite{pocher:napac22-tuze3}. This ML technique in comparison to our SINDy fit introduced over three orders of magnitude more coupled equations\footnote{That ML technique did not use differential equations, but rather a discrete, nonlinear map iterating time steps. The distinction between this and the discretized system of equations is subtle, yet important to specify.}, in addition to additional ML-associated hyperparameters. Furthermore, no physics can be interpreted from the ML model, in contrast to the directly interpretable SINDy fits whose coefficients we used to validate the return time of the SHO term. For that reason, we believe our SINDy technique, especially after being used with our synthetically imposed impediments, offers a compelling alternative for predicting nonlinear beam dynamics from ML techniques. 
 
On the other hand, the centroid model we deployed obtained accurate qualitative and statistical agreement with the Warp simulation data over many turns, \cref{fig:xc_yc_mod_fourier}. This simple model outperformed both the ML technique performed by our colleagues or the SINDy fits described within this paper, \cref{fig:phase_all_comp,fig:fourier_all}. The advantages of having this analytic model were twofold. One, we had explicit expressions for $x_c''$ and $y_c''$ with which we can interpret the SINDy results that were obtained without recourse to the model, and in consequence contained terms that were nonphysical, see App.~\ref{sec:app:Fourier-data}. Two, we were able to interpret our proposed FOM, the power spectra of $x_c$ and $y_c$, with high fidelity and can explain its fundamental structure. While we accede the capability of this first principles modeling for centroid dynamics, we must acknowledge the fact and prepare for the inevitability that not every accelerator lattice has neat, closed-form expressions for every element that can be used in such simple, low-dimensional differential equations. Higher-order nonlinearities, misalignments, instabilities, etc. all contribute to dynamics not present in first-principle models.

In addition to being directly interpretable, the SINDy fits obtain an order of magnitude computational acceleration when compared to Warp simulation. The Warp simulations, for the particular numerical settings within this paper, took 15 seconds to run one UMER turn on a Linux machine (Warp is compiled in Fortran and C and run from a Python wrapper), while the effectively two-dimensional SINDy equations ran the same turn in approximately 1 second with a simple direct integration technique in Python. With numerical optimization and running with a compiled code, this computational speed up would only increase.

An example extension of this work is to apply our results to the real experiment on UMER, and other machines. Instead of taking measurements every 5~mm, a luxury not realizable in physical storage rings, we could modify our approach to take Poincar\'e sections at discrete intervals. UMER has 12 beam position monitors we could use to generate configuration space data. We would then modulate SINDy to change from learning a continuous differential equation along the lattice to learning a discrete dynamical map for the beam's centroid. For beam lifetimes of approximately 100 turns, these collection points can provide thousands of configuration space data slices, a similar data amount to what was used in this paper.

In summary, we believe SINDy is a promising method that can hasten accelerator commissioning by uncovering underlying beam dynamics. We have shown both recoverable beam dynamics \cref{fig:SINDy_param}(c) and FOMs \cref{fig:SINDy_fourier}(c) in a manner that is interpretable through both having analytic terms corresponding to specific physics, and also to the centroid model. With our methodology of performing careful data analysis combined with intuition of physical forcing functions, we aim to develop compact beam dynamics models with high fidelity. This is applicable in real accelerators where purely analytic models don't exist or where forcing functions are unknown.

\begin{acknowledgments}
We would like to thank Levon Dovlotyan for collaborative discussion, and Daniel P. Lathrop for both collaborative discussion and editing. We would also like to thank the SINDy community for enabling this work. This work has been supported by US DOE-HEP grant DE-SC0022009.
\end{acknowledgments}

\appendix

\section{\label{sec:app_centroid}Beam Moment Evolution Equations} 

Moments of particle beams are taken with respect to every particle in the underlying distribution. Here, we take angle brackets with a perpendicular subscript to denote an average over the four dimensional trace space\footnote{In accelerator physics, trace space is equivalent to the more traditional phase space but with a dimensionless momentum variable.} $(x,y,x',y')$ with the beam's transverse distribution function $f_{\perp} = f_{\perp}(x,y,x',y',z)$
\begin{equation*}
   \langle ... \rangle_{\perp} = \frac{\int d^2 x_{\perp} \int d^2 x_{\perp}' ... f_{\perp}}{\int d^2 x_{\perp} \int d^2 x_{\perp}' f_{\perp}}
\end{equation*}
with $...$ a stand in quantity within the angle brackets, $x_{\perp}$ and $x_{\perp}'$ transverse trace space coordinates, and $\int d^2 x_{\perp} \int d^2 x_{\perp}' f_{\perp}$ a normalization condition.

Two equations sets used in beam physics for lattice design and optimization are the envelope equations and the centroid equations. The envelope equations describe the evolution of the root-mean-square (rms) beam radius, while the centroid equations describe the evolution of mean displacement. We use terms from the envelope equation to yield dimensionless parameters, while analyzing the centroid equations in this paper. 

The envelope $\sigma_x^2 = \langle x^2 \rangle_{\perp}$ is a measure of rms beam edge. The envelope equation describes how applied focusing forces, repulsive space charge (self-repelling electric force), and thermal forces (pressure) interact to evolve beam size along the lattice. A useful dimensionless beam parameter is the ratio of space charge to thermal forces $K \sigma_x^2 \varepsilon_x^{-2}$~\cite{reiser2008theory}. The quantity $\varepsilon$ is the rms emittance, a statistical measure of beam area in trace (phase) space\footnote{Phase space is the space spanned by all the particles positions $\mathbf{x}$ and momenta $\mathbf{p}$. Trace space is directly related to phase space, but instead of using particle momenta as the conjugate coordinate, makes the momentum coordinate dimensionless by dividing by the axial beam momentum $p_z=\gamma m v_z$.} calculated with averages over the whole beam's transverse trace space. The quantity $K$ is the beam's generalized perveance, a dimensionless measure of electrostatic self-repulsion effects to inertial effects in the beam and is defined as
$K=2 I \,  I_0^{-1} \beta^{-3} \gamma^{-3}$
with: $I_0$ a characteristic current ($I_0 = 4 \pi \varepsilon_0 m c^3 q^{-1}$) which is 17~kA for electrons with $\varepsilon_0$ being the permittivity of free space, $m$ electron mass, $q$ the elemental charge; and $I$ beam current. 

The general centroid equations take the form
\begin{subequations}
\begin{align} 
    \label{eqn:xc_cent_gen} x_c'' + \frac{\gamma' \beta'}{\gamma \beta} x_c + \underbrace{\left[ \kappa_{x} (z) + \frac{1}{R_x^2(z)} \right]}_{\text{lattice forces}} x_c &= \underbrace{ \frac{2 \pi \varepsilon_0 K }{\lambda} \left\langle E_{x}^{(i)} \right\rangle_{\perp} }_{\text{pipe image forces}}  \, , 
    \\ 
    \label{eqn:yc_cent_gen} y_c'' + \frac{\gamma' \beta'}{\gamma \beta} y_c +  \underbrace{\left[ \kappa_{y} (z) + \frac{1}{R_y^2(z)} \right]}_{\text{lattice forces}} \, y_c &=  \underbrace{\frac{2 \pi \varepsilon_0 K}{\lambda} \left\langle E_{y}^{(i)} \right\rangle_{\perp}}_{\text{pipe image forces}} \, .
\end{align}
\end{subequations}
\Cref{eqn:xc_cent_gen,eqn:yc_cent_gen} have no space charge forces, which average to zero under $\langle ... \rangle_{\perp}$.

As also described in Sec.~\ref{sec:method:UMER_lat}, $x_c / R_x^2(z)$ represents the change in the balance between the bending force of the dipoles and the centripetal acceleration implied by the curved trajectory.  If the beam is displaced radially such that $x_c > 0$, the centripetal acceleration decreases and the dipole force exceeds the centripetal acceleration, causing the beam to be directed back toward the reference trajectory.  Similarly, if the beam is radially displaced inward such that $x_c < 0$, the increased centripetal acceleration causes the beam to be directed outward toward the reference trajectory. Although this term represents a competition between two effects, centripetal acceleration, and magnetic dipole force, it is represented in terms of a single variable related to an effective radius of curvature.

Additionally, $1/R_x(z) = B_{y}^{(a)}(z)[B \rho]^{-1}$ being an effective bend radius in the $\hat{x}$ direction accounting for the average centripetal acceleration and the dipole magnetic field \footnote{We assume a sector dipole model in both the Warp simulation and the centroid model. In this model, the accelerator coordinates transform to a Frenet–Serret coordinate system describing a bending beam. The effective force $x_c/R_x^2(z)$ approximates both the dipole force on the beam and the centripetal acceleration caused by the bending.} $B_{y}^{(a)}$ where the superscript $(a)$ means applied and $y$ is the orthogonal transverse coordinate to $x$, and $[B \rho] = p_0 q^{-1} = \gamma m \beta c q^{-1}$ is the magnetic rigidity.  The same treatment may be applied to the term $y_c/R_y^2(z)$. 

As also introduced in Sec.~\ref{sec:method:UMER_lat}, the terms $\kappa_x$ and $\kappa_y$ describe the effect of the quadrupoles.  If, as assumed here, the quadrupoles are properly aligned with respect to the normal to the plane of the ring, the quadrupole force is in the same direction $(+/-)$ as the displacement.  Further, if a quadrupole is focusing (defocusing) in the $\hat{x}$-direction, it will be defocusing (focusing) in the $\hat{y}$-direction.  As a result, $\kappa_x = - \kappa_y$.  The strength of the quadrupole field is expressed as $\kappa_x = qG(z)(m \gamma \beta c)^-1$~\cite{wiedemann2015particle}. Here $q$ is the particle charge,
$G(z)$ is the spatially dependent quadrupole gradient in units of T/m, $\beta = v_z/c$, $v_z$ is the beam velocity, $c$ is the speed of light, and $\gamma = (1 - \beta^2)^{-1/2}$ is the relativistic factor. Both of the lattice forces, quadrupole and dipole) are functions of distance $z$ along the reference trajectory. The quadrupole force profile is given by $G(z)$. The dipole-centripetal force is a function of distance $z$ along the reference trajectory, given by the profile function $1/R^2(z)$. The functional form of these profiles depends on details of the dipole and quadrupole fields. The profile dependence is often treated approximately as a series of hard-edge functions, and that is what we do here.

The terms that appear on the right side of \cref{eqn:xc_cent_gen,eqn:yc_cent_gen} represent the effect of the self field of the charged particles in the beam.  There is no net force on the centroid of the beam due to direct particle interactions through Coulomb's law. However, the beam is almost always surrounded by an electrically conducting tube, and the image charges in the walls of this tube attract the charges in the beam. This results in a force on the centroid that is in the same direction and same sign as the displacement. This force may be estimated by treating the beam as a single particle inside a conducting tube, and is treated in various textbooks~\cite{jackson1999classical,reiser2008theory}. The image electric field produced by the pipe (e.g. $E_x^(i)$) is averaged over the entire beam to find the average electric field felt by the beam. The additional terms on the right-hand side: $\lambda$ the axial line charge density $(\lambda = I \beta^{-1} c^{-1})$; and $\left\langle E_{x}^{(i)} \right\rangle_{\perp}$ a transverse average of the pipe's image forces in the $\hat{x}$-direction induced by the beam.

The induced image charge force to the lowest order is linear in the centroid variable~\cite{lee:pac87}. \Cref{eqn:xc_cent_gen,eqn:yc_cent_gen} are simplified to the following for constant axial beam velocity ($\gamma'=\beta'=0$) and with the assumption that centroid deviations are much smaller than the pipe radius that produces image forces,
\begin{align} 
    \label{eqn:xc_cent_simp} x_c'' + \left[\kappa_{x}(z) + \frac{1}{R^2(z)} - \frac{K}{r_p^2} \right] x_c &= 0 \, , 
    \\ 
    \label{eqn:yc_cent_simp} y_c'' + \left[ - \kappa_{x}(z) - \frac{K}{r_p^2} \right] y_c &= 0
\end{align}
The quantity $\kappa_{x}(z) =  q \, G(z)(m\gamma \beta c)^{-1}$ where $G(z)$ is the quadrupole gradient with units of T/m. We use the same quadrupole gradients in the centroid model and Warp simulation up to linear order. 

We establish the size of the image charge force---$x_c K/r_p^2$, the rightmost term in brackets in \cref{eqn:xc_cent_simp}---by comparing dimensionless ratios of it with the other terms in brackets in \cref{eqn:xc_cent_simp}. We take the average of the ratio of quadrupole focusing---$\kappa_x(z) \, x_c $, the leftmost term in brackets in \cref{eqn:xc_cent_simp}---to the image charge force
\begin{equation*}
    \frac{1}{L}\int_{0}^{L} dz \, \kappa_{x} (z) \frac{r_p^2}{K} 
\end{equation*} 
over one turn of length $L$ in the superperiod of UMER. The resultant calculation yields approximately 3000, a three-order-of-magnitude difference in quadrupole lattice forces\footnote{The quadrupole and dipole forces are comparable to each other.} to the pipe's image charge forces. Leveraging this, we write the two equations for $x_c$ and $y_c$ from \cref{eqn:xc_cent_simp,eqn:yc_cent_simp} as
\begin{align} 
    \label{eqn:xc_cent_app} x_c'' + \left[ \kappa_{x}(z) - \frac{1}{R^2(z)} \right] x_c &= 0 \, , 
    \\ 
    \label{eqn:yc_cent_app} y_c''  - \kappa_{x}(z) y_c &= 0
\end{align}
which are decoupled aside from $\kappa_x = - \kappa_y$ due to quadrupole field structure. The term $\kappa_x$ in \cref{eqn:yc_cent_app} is positive so that it is defocusing for $x$ when the beam passes through the first quadrupole. \Cref{eqn:xc_cent_app,eqn:yc_cent_app} are the equations solved in this paper and are reproduced---with the inclusion of $\kappa_y$---in \cref{eqn:xc_cent,eqn:yc_cent}.

\section{\label{sec:app_Coef_Tab}Tables of Model Coefficients}

This appendix reports the various coefficients for all the models in Sec.~\ref{sec:res} found using the STLSQ optimizer. 

We calculate the natural linear frequency replicating a SHO by multiplying the entries for $y_c$ in \cref{tab:model_coefs_xc} with the entries for $x_c$ in entries in \cref{tab:model_coefs_yc} and dividing $2\pi$ by the square root $k_N/ 2 \pi \approx  \left(-\xi_{x_c}^{(y_c)} \xi_{y_c}^{(x_c)} \right)^{-1/2} $. For fit 2, fit 3, and case 3 the $k_N/2\pi$ are 0.58~m$^{-1}$, 0.63~m$^{-1}$, and 0.47~m$^{-1}$, respectively, which compare favorably with the measured value of 0.61~m$^{-1}$. 

\begin{table}[b]
	\setlength\tabcolsep{3.5pt}
	\centering
	\caption{\label{tab:model_coefs_xc}Coefficients for the various $d x_c/ dz$ models. Fit 3, \cref{fig:SINDy_param}(c), is also case 1 shown in \cref{fig:case_comp}(a).}
 \begin{ruledtabular}
	\begin{tabular}{lccccr}
		Term & fit 1 & fit 2 & fit 3 & case 2 & case 3 \\ \hline
		1.0 & -2.29e-4 & 0.0 & 6.30e-5 & 0.0 & 0.0 \\ 
		$z$ & n/a & 0.0 & 1.54e-5 & 0.0 & 0.0 \\
		$x_c$ & n/a & 0.0 & -0.490 & 0.0 & 0.0  \\
            $y_c$ & n/a & 4.45 & 5.91 & 0.0 & 3.76   \\
            cos$(k_0 z)$ & 5.32e-3 & -1.24e-3 & -1.84e-3 & 5.32e-3 & -1.84e-5  \\
            cos$(k_1 z)$ & 4.41e-4 & 0.0 & -2.29e-3 & 0.0 & -2.67e-3  \\
            cos$(k_2 z)$ & 5.80e-5 & 0.0 & -2.84e-3 & 0.0 & -2.70e-3  \\
            sin$(k_0 z)$ & 3.79e-4 & 0.0 & 3.51e-3  & 0.0 & 2.83e-6 \\
            sin$(k_1 z)$ & 4.14e-3 & 2.98e-3 & 2.97e-3 & 4.14e-3 & 3.14e-3  \\
            sin$(k_2 z)$ & -2.91e-3 & -3.43e-3 & -3.55e-3 & -2.91e-3 & -3.26e-3  \\
            $z$ cos$(k_0 z)$ & n/a & n/a & 2.82e-4 & 0.0 & 0.0 \\
            $z$ cos$(k_1 z)$ & n/a & n/a & 4.79e-4 & 0.0 & 5.57e-4  \\
            $z$ cos$(k_2 z)$ & n/a & n/a & 5.01e-4 & 0.0 & 4.69e-4  \\
            $z$ sin$(k_0 z)$ & n/a & n/a & -2.58e-4 & 0.0 & 1.67e-4  \\
            $z$ sin$(k_1 z)$ & n/a & n/a & -7.07e-5 & 0.0 & 0.0  \\
            $z$ sin$(k_2 z)$ & n/a & n/a & 3.24e-6 & 0.0 & 0.0 \\
            $x_c$ cos$(k_0 z)$ & n/a & n/a & 7.79e-2 & 0.0 & 3.45e-2  \\
            $x_c$ cos$(k_1 z)$ & n/a & n/a & 6.06e-2 & 0.0 & 7.08e-2 \\
            $x_c$ cos$(k_2 z)$ & n/a & n/a & 8.75e-2 & 0.0 & 8.35e-2  \\
            $x_c$ sin$(k_0 z)$ & n/a & n/a & -0.105 & 0.0 & -2.10e-2  \\
            $x_c$ sin$(k_1 z)$ & n/a & n/a & 4.60e-2 & 0.0 & 0.0 \\
            $x_c$ sin$(k_2 z)$ & n/a & n/a & -1.40e-2 & 0.0 & -4.22e-2  \\
            $y_c$ cos$(k_0 z)$ & n/a & n/a & -7.27e-2 & 0.0 & 4.60e-2  \\
            $y_c$ cos$(k_1 z)$ & n/a & n/a & 5.47e-3 & 0.0 & 1.35e-2 \\
            $y_c$ cos$(k_2 z)$ & n/a & n/a & -1.88e-2 & 0.0 & -4.20e-2  \\
            $y_c$ sin$(k_0 z)$ & n/a & n/a & -4.04e-2 & 0.0 & 2.46e-2  \\
            $y_c$ sin$(k_1 z)$ & n/a & n/a & 0.121 & 0.0 & 0.118  \\
            $y_c$ sin$(k_2 z)$ & n/a & n/a & -9.71e-2 & 0.0 & -0.101 \\
	\end{tabular}
  \end{ruledtabular}
\end{table}

Different optimization parameters yield different coefficients for our prescribed basis function in \cref{eqn:fi}. A trend that emerges is a cascading order of coefficients. In decreasing order, they are: the SHO terms $\xi_c$, the nonlinear $\xi_c \, \text{cos}(k_i z)$ terms, the linear $\text{cos}(k_i z)$ terms, the nonlinear $z \, \text{cos}(k_i z)$ terms, the linear $z$, and constants. This represents the dominance of the SHO in projected dynamics, over that of lattice forcing, in the centroid model.

\begin{table}[b]
	\setlength\tabcolsep{3.5pt}
	\centering
	\caption{\label{tab:model_coefs_yc}Coefficients for the various models for $d y_c/ dz$. Fit 3, \cref{fig:SINDy_param}(c), is also case 1 shown in \cref{fig:case_comp}(a).}
 \begin{ruledtabular}
	\begin{tabular}{lccccr}
		Term & fit 1 & fit 2 & fit 3 & case 2 & case 3 \\ \hline
		1.0 & 2.89e-5 & 0.0 & -3.44e-4 & 0.0 & 0.0 \\ 
		$z$ & n/a & 0.0 &  3.41e-5 & 0.0 & 0.0  \\
		$x_c$ & n/a & -2.94 & -2.62 & 0.0 & -2.36   \\
            $y_c$ & n/a & 0.0 & 0.826 & 0.0 & -0.390   \\
            cos$(k_0 z)$ & -5.59e-4 & 0.0 & -2.43e-3 & 0.0 & -1.07e-3 \\
            cos$(k_1 z)$ & 4.15e-3 & 3.38e-3 & 3.38e-3 & 4.15e-3 & 3.57e-3 \\
            cos$(k_2 z)$ & 2.76e-3 & 3.13e-3 & 2.95e-3 & 2.76e-3 & 3.12e-3 \\
            sin$(k_0 z)$ & -5.62e-3 & -1.17e-3 & 1.34e-5 & -5.62e-3 & -2.09e-3  \\
            sin$(k_1 z)$ & 3.75e-4 & 0.0 & 3.04e-3 & 0.0 & 3.37e-3 \\
            sin$(k_2 z)$ & -3.59e-4 & 0.0 & -2.66e-3  & 0.0 & -2.52e-3 \\
            $z$ cos$(k_0 z)$ & n/a & n/a & 5.80e-5 & 0.0 & 1.42e-4 \\
            $z$ cos$(k_1 z)$ & n/a & n/a & 2.85e-5 & 0.0 & 0.0 \\
            $z$ cos$(k_2 z)$ & n/a & n/a & 3.25e-5 & 0.0 & 0.0 \\
            $z$ sin$(k_0 z)$ & n/a & n/a & -2.64e-4 & 0.0 & 0.0 \\
            $z$ sin$(k_1 z)$ & n/a & n/a & -4.72e-4 & 0.0 & -4.76e-4  \\
            $z$ sin$(k_2 z)$ & n/a & n/a & 3.95e-4 & 0.0 & 3.94e-4 \\
            $x_c$ cos$(k_0 z)$ & n/a & n/a & 0.184 & 0.0 & 2.25e-2 \\
            $x_c$ cos$(k_1 z)$ & n/a & n/a & 4.51e-2 & 0.0 & 0.0 \\
            $x_c$ cos$(k_2 z)$ & n/a & n/a & 2.66e-2 & 0.0 & 2.89e-2 \\
            $x_c$ sin$(k_0 z)$ & n/a & n/a & 0.123 & 0.0 & 2.67e-2 \\
            $x_c$ sin$(k_1 z)$ & n/a & n/a & 3.79e-2 & 0.0 & 2.37e-3 \\
            $x_c$ sin$(k_2 z)$ & n/a & n/a & -5.41e-3 & 0.0 & -2.21e-2 \\
            $y_c$ cos$(k_0 z)$ & n/a & n/a & 8.79e-2 & 0.0 & -2.04e-2 \\
            $y_c$ cos$(k_1 z)$ & n/a & n/a & 4.38e-2 & 0.0 & 0.0 \\
            $y_c$ cos$(k_2 z)$ & n/a & n/a & -2.23e-2 & 0.0 & -1.35-2 \\
            $y_c$ sin$(k_0 z)$ & n/a & n/a & -0.165 & 0.0 & -2.00e-2 \\
            $y_c$ sin$(k_1 z)$ & n/a & n/a & 2.39e-2 & 0.0 & 4.69e-2 \\
            $y_c$ sin$(k_2 z)$ & n/a & n/a & -3.60e-2 & 0.0 & -3.28e-2 \\
	\end{tabular}
  \end{ruledtabular}
\end{table}

\section{\label{sec:app_dipo}Dipole Parameters}

The hard-edge sector dipoles we describe in Sec.~\ref{sec:centroid} have strengths that are periodic on UMER's superperiod. \Cref{tab:model_dipole_by} enumerates the index $i$, start location $z_s$, end locations $z_e$, and strengths $B_y$ of each of the 36 dipoles in UMER. These dipoles bend the beam discretely, and create anticorrelated motion we observe in the data and the centroid model on the five turn timescale. Eventually, this anticorrelated motion secularly precesses around the beam center on the order of 200~m or $\sim20$ turns, which results in the reflection symmetry about the horizontal and vertical axes in \cref{fig:phase_all_comp}(a) and \cref{fig:phase_all_comp}(c).

\begin{table}[]
\setlength\tabcolsep{3.5pt}
\centering
\caption{\label{tab:model_dipole_by}Magnetic field strengths $B_y$ and locations $z \in (z_{\text{start},i},z_{\text{end},i})$ of each dipole $i$ around the UMER lattice, bending the beam in the $\hat{x}$-direction. Even though the simulation does not include the Earth's field, the variations in dipole strengths reflect the fact that the real UMER experiment account for local variations in the Earth's field.}
\begin{ruledtabular}
\begin{tabular}{lccr}
$i$ & $z_{\text{start}}$ [m] & $z_{\text{end}}$ [m]& $B_y$ [Gs] \\ \hline
1  & 0.142  & 0.178  & -13.58 \\
2  & 0.462  & 0.498  & -14.98 \\
3  & 0.782  & 0.818  & -13.75 \\
4  & 1.102  & 1.138  & -14.32 \\
5  & 1.422  & 1.458  & -14.88 \\
6  & 1.742  & 1.778  & -14.78 \\
7  & 2.062  & 2.098  & -14.70 \\
8  & 2.382  & 2.418  & -14.64 \\
9  & 2.702  & 2.738  & -14.45 \\
10 & 3.022  & 3.058  & -13.92 \\
11 & 3.342  & 3.378  & -13.70 \\
12 & 3.662  & 3.698  & -13.48 \\
13 & 3.982  & 4.018  & -13.37 \\
14 & 4.302  & 4.338  & -13.27 \\
15 & 4.622  & 4.658  & -13.11 \\
16 & 4.942  & 4.978  & -12.91 \\
17 & 5.262  & 5.298  & -12.67 \\
18 & 5.582  & 5.618  & -12.64 \\
19 & 5.902  & 5.938  & -12.85 \\
20 & 6.222  & 6.258  & -14.06 \\
21 & 6.542  & 6.578  & -13.48 \\
22 & 6.862  & 6.898  & -13.64 \\
23 & 7.182  & 7.218  & -13.36 \\
24 & 7.502  & 7.538  & -12.86 \\
25 & 7.822  & 7.858  & -12.52 \\
26 & 8.142  & 8.178  & -13.53 \\
27 & 8.462  & 8.498  & -13.37 \\
28 & 8.782  & 8.818  & -13.96 \\
29 & 9.102  & 9.138  & -13.97 \\
30 & 9.422  & 9.458  & -13.34 \\
31 & 9.742  & 9.778  & -13.42 \\
32 & 10.062 & 10.098 & -13.49 \\
33 & 10.382 & 10.418 & -12.61 \\
34 & 10.702 & 10.738 & -13.04 \\
35 & 11.022 & 11.058 & -9.83  \\
36 & 11.342 & 11.378 & -14.30
\end{tabular}
\end{ruledtabular}
\end{table}

\section{\label{sec:app:Fourier-data}Fourier Transforms Analysis}

We take the Fourier transform as the integral 
\begin{equation*}
    \label{eqn:app:fourier:fourier}
    \mathcal{F}[f(z)] = \hat{f}(k) = \frac{1}{\sqrt{2 \pi}} \int_{-\infty}^{\infty} dz \,  e^{i k z} \, f(z)
\end{equation*}
that integrates a signal $f(z)$ over all $z$, transforming to the reciprocal domain $k$. To gain understanding of how the power spectra in \cref{fig:Centroid_Fourier}(a) and \cref{fig:xc_yc_mod_fourier} arose, we solved two differential equations and obtained their power spectra. We compare these power spectra to analytic terms arising from linear and nonlinear interactions.

The forced, harmonic oscillator is
\begin{equation} \label{eqn:app:fourier:lin}
    x'' = \text{cos}(b z) - a^2 x
\end{equation}
where $x$ is the dependent variable, $z$ the independent variable, $a$ the natural angular frequency of the system, and $b$ the forcing angular frequency. The general solution is
\begin{align}
\nonumber
    x(z) = &\left(x_0 +  \frac{1}{a^2 - b^2} \right) \text{cos}(b z) - \\  &\frac{1}{a^2 - b^2} \text{cos}(az) + \frac{x_0'}{b} \text{sin}(bz) \, .
    \label{eqn:app:fourier:linsol}
\end{align}
which has the following infinite time limit Fourier transform $\mathcal{F}[...]$ with $x_0=x_0'=0$ for simplicity 
\begin{align*}
    \mathcal{F}\left[ x(z) \right] = \frac{\sqrt{\pi/2}}{a^2 - b^2} \left[ \delta(k \pm a) + \delta(k \pm b) \right] \, .
\end{align*}

A similar forced, second-order differential equation is 
\begin{equation*}
\label{eqn:app:fourier:nonlin}
    x'' = x \,\text{cos}(b z) - a^2 x
\end{equation*}
with the only modification being a nonlinear interaction between $x$ and the periodic forcing term $\text{cos}(b z)$. This equation is a low order approximation to \cref{eqn:xc_cent_simp,eqn:yc_cent_simp}, and doesn't have an analytic solution, preventing an analytic Fourier transformation.

To obtain a direct comparison of the power spectra for solutions of these equations, we numerically integrated them 100 dimensionless time units with a time step of $dz=0.01$. We set the natural frequency of the system, inverse of the return time, $a=1$, and the forcing frequency $b=2\pi$. 

\begin{figure}[!htbp]
  \centering
  \includegraphics*[width=\columnwidth]{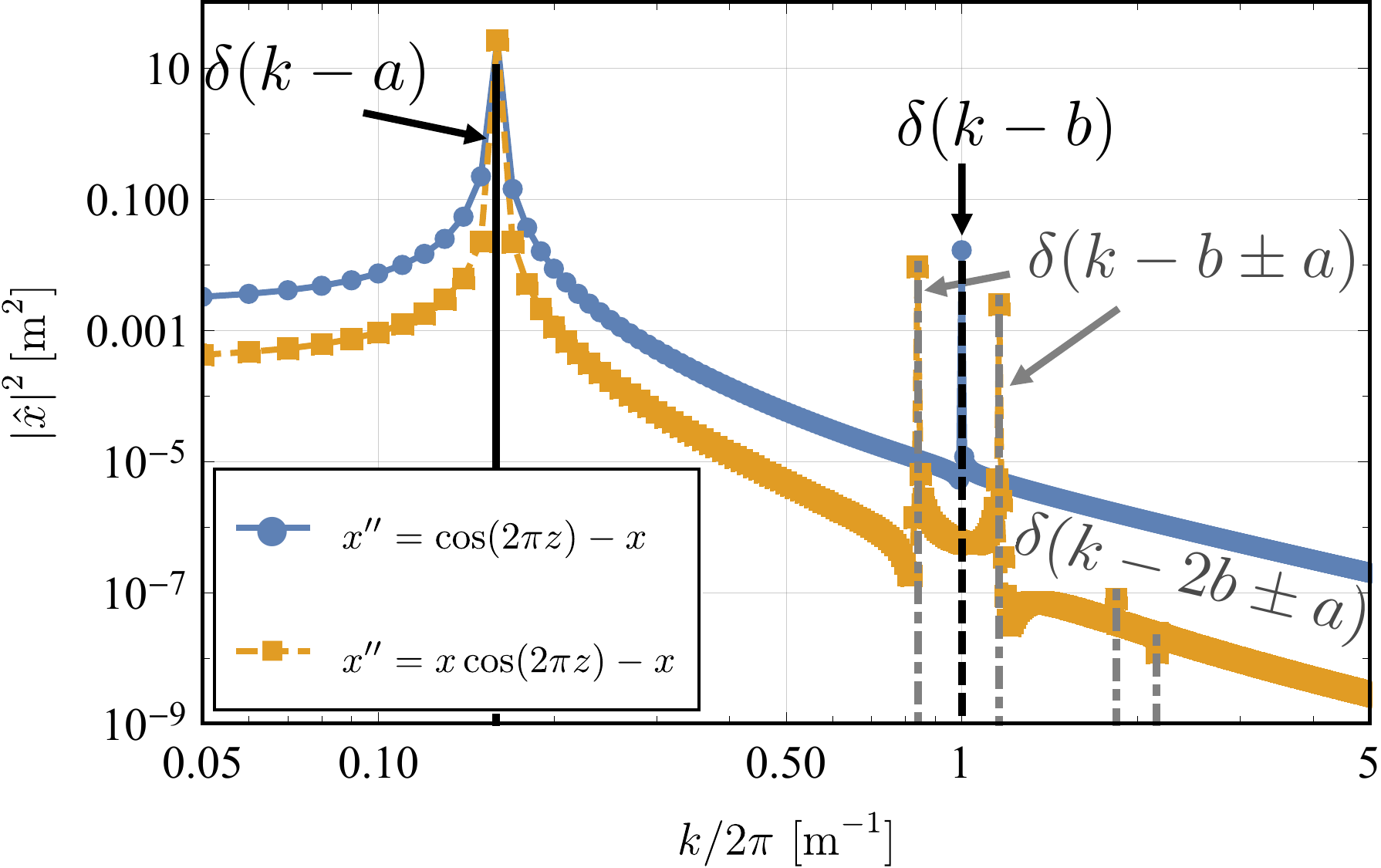}
\caption{\label{fig:discrete_PS}Unregularaized power spectra for the solution to differential equations $x''= \text{cos}(2\pi z) - x$ in blue circles and $x''= x \, \text{cos}(2\pi z) - x$ in yellow squares. The solid black line is $k=a$, the natural frequency of the system, the dashed black line is $k=b$, while the dash-dot grey lines are $k/2\pi = n \, b \pm a$ with $n=1$ or $2$.}
\end{figure} 

Calculating the power spectra of the solution for these forced harmonic oscillators yields \cref{fig:discrete_PS}, with the familiar peak pair as seen in \cref{fig:Centroid_Fourier}(a). These peaks represent frequency mixing, which we can replicate analytically by taking Fourier transforms of $\mathcal{F}\left[ \text{cos}(n b z) \text{cos}(a z) \right]$ with integer $n$. Other structures observable in \cref{fig:Centroid_Fourier}(a) include various harmonics of the natural frequency, and in addition, higher order, though lower amplitude, behavior from  $\text{cos}(n b z)^2$ terms interacting with the natural frequency, perhaps caused by the purely negative dipole fields.

Taking a finite time Fourier transform from 0 to $L$ of $\text{cos}(a z)$ and calculating its power spectra, we obtain
\begin{align*}
    & \left|\mathcal{F}[x(z)]\right|^2 =\frac{1}{(k + a)^2} \left[ 1 - \text{cos}[L(a+k)] \right] \\
    & - \frac{1}{(a+k)(a-k)} \left[ 1 + \text{cos}(2aL) + \text{cos}(L a) \text{cos}(L k) \right] \\
    & + \frac{1}{(a - k)^2} \left\{ 1 - \text{cos}[L(a-k)] \right\}
\end{align*}
which are Lorenztians centered at $k = \pm a$. Extending this to \cref{eqn:app:fourier:linsol}, we note the peaks at $k=a$ and $k=b$ are superimposed Lorenztians for blue circles in \cref{fig:discrete_PS}. We also note the $1/k^2$ falloff observable in all the power spectra in the plots present comes from this finite length integral. The finite integration bounds and cosine arguments modulating the Lorenztians provide periodic behavior to the amplitude of the $1/k^2$ falloff, which is not observable for numerical power spectra.

We regularized the centroid data in \cref{fig:UMER_lat_phase}(b) by multiplying it by a function $\varphi(z)$
\[
\varphi(z) = \begin{cases} 
      \text{sin}^2\left(k_s \dfrac{z}{L}\right) & z\leq
      \frac{\pi}{2} \frac{L}{k_s}\\
     1 & \frac{\pi}{2} \frac{L}{k_s} < z < L \left(1 - \frac{\pi}{2 k_s} \right)  \\
      \text{sin}^2\left(k_s \dfrac{z}{L}\right) & L \left(1 - \frac{\pi}{2 k_s} \right) \leq z
   \end{cases}
\]
with $k$ the angular frequency, and $L$ the length of the regularized time series. We took the Fourier transform of the total signal $g(z) = \varphi(z) f(z)$, which induced a convolution in frequency space~\cite{arfken2011mathematical}
\begin{align*}
    \mathcal{F}[g(z)] = \hat{g}(k) &= \frac{1}{\sqrt{2 \pi}} \int_{-\infty}^{\infty} dz \,  e^{i k z} \, \varphi(z) \, f(z) \\
    & = \int_{- \infty}^{\infty} dk' \varphi(k') f(k - k')
\end{align*}
The convolution length scale in frequency space is set by the $k_s$ in $\varphi(z)$; higher $k_z$ have broader envelopes for $\hat{\varphi}(k)$, but faster rise times in $z$. We set $k_s=2\pi$, which, although broad in length space, is narrow in frequency space, inducing significant signal falloff, allowing us to observe spectra structure far below the unregularized signal. 

\begin{figure}[!htbp]
  \centering
  \includegraphics*[width=\columnwidth]{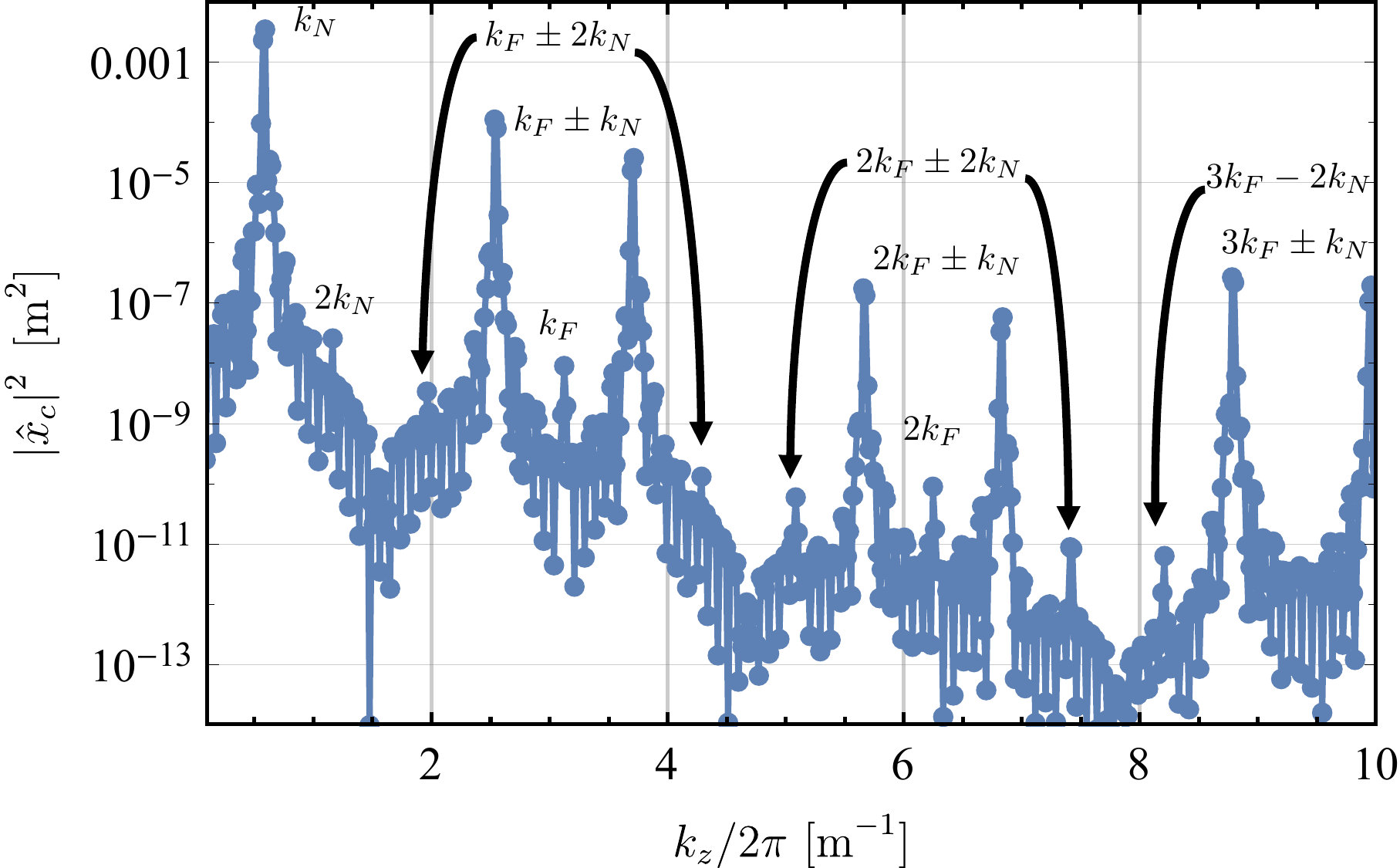}
 \caption{\label{fig:xc_fourier_turn}Semilog plot of the regularized power spectrum of $|\hat{x}_c|^2$ over five UMER lattice turns. Distinctive structures at relevant integer multiples of constituent frequencies in the power spectra are marked.}
\end{figure} 

\Cref{fig:xc_fourier_turn} shows the regularized power spectra for $\hat{x}_c(k_z)$. These regularized power spectra reveal a structure that we now confidently analyze in the context of both the centroid model described in Sec. \ref{sec:centroid} and the Fourier analysis of various differential equations described in this appendix. Two harmonics of the natural frequency $k_N = 0.61$~m$^{-1}$ are noted in this plot at 0.61~m$^{-1}$ and 1.21~m$^{-1}$. Further harmonics of $k_N$ may be present in \cref{fig:xc_fourier_turn}, but we cannot definitely distinguish them from the power spectrum's numerical floor. We also note frequency mixing harmonics, with peaks observed at multiple integer multiples of the forcing frequency mode-mixing with the natural frequency's first and second harmonics. 

\bibliography{physrevE}

\end{document}